\documentclass[a4paper]{article}

\usepackage{color}
\usepackage{INTERSPEECH2020}
\usepackage{url}
\usepackage{multicol}
\usepackage{multirow}
\usepackage{graphicx}
\usepackage{rotating}
\usepackage{cite,balance}

\title{Voice Conversion Challenge 2020 \\ –- Intra-lingual semi-parallel and cross-lingual voice conversion –-}
\name{Zhao Yi$^{1}$$^{\star}$\thanks{$\star$ Equal contribution.}, Wen-Chin Huang$^{2}$$^{\star}$, Xiaohai Tian$^{3}$$^{\star}$, Junichi Yamagishi$^1$$^{\star}$, \\ Rohan Kumar Das$^3$,  Tomi Kinnunen$^4$, Zhenhua Ling$^5$, Tomoki Toda$^2$}
\address{
  $^1$National Institute of Informatics, Japan
  $^2$Nagoya University, Japan\\
  $^3$National University of Singapore, Singapore
  $^4$University of Eastern Finland, Finland \\ 
  $^5$University of Science and Technology of China, P.R.China
}
\email{vcc2020@vc-challenge.org}

\begin{document}

\maketitle
\begin{abstract}
The voice conversion challenge is a bi-annual scientific event held to compare and understand different voice conversion (VC) systems built on a common dataset. In 2020, we organized the third edition of the challenge and constructed and distributed a new database for two tasks, intra-lingual semi-parallel and cross-lingual VC. After a two-month challenge period, we received 33 submissions, including 3 baselines built on the database. From the results of crowd-sourced listening tests, we observed that VC methods have progressed rapidly thanks to advanced deep learning methods. In particular, speaker similarity scores of several systems turned out to be as high as target speakers in the intra-lingual semi-parallel VC task. However, we confirmed that none of them have achieved human-level naturalness yet for the same task. The cross-lingual conversion task is, as expected, a more difficult task, and the overall naturalness and similarity scores were lower than those for the intra-lingual conversion task. However, we observed encouraging results, and the MOS scores of the best systems were higher than 4.0. We also show a few additional analysis results to aid in understanding cross-lingual VC better.
\end{abstract}
\vspace{1mm}

\section{Introduction}

Voice conversion (VC) is a technique for transforming non-/para-linguistic information, such as speaker identity, included in a given speech waveform into the desired information while preserving the linguistic information of the speech waveform. VC offers great potential for the development of various new applications, such as a speaking aid for vocally handicapped people \cite{Kain2007dysarthric}, computer-assisted language learning leveraging accent conversion \cite{Felps2009accent}, a voice changer for generating various types of expressive speech \cite{Turk2010expressive}, a novel vocal effector for singing voices \cite{Villavicencio2010singing}, and enhanced telecommunication with silent speech interfaces \cite{Toda2012silent}. Thanks to a well-formulated approach to describing VC as a regression problem, VC research has been popularized through the vast sharing of various data-driven techniques. However, comparing across several VC techniques is not straightforward as their performance strongly depends on the speech datasets used by individual researchers.

The Voice Conversion Challenge (VCC) was launched in 2016 to better understand different VC techniques by comparing their performance using a freely available dataset as a common dataset and by bringing together different teams to look at a common goal and to share views about unsolved problems and challenges faced by the current VC techniques \cite{toda2016voice}. In the previous VCCs held in 2016 \cite{toda2016voice} and 2018 \cite{Lorenzo-Trueba2018}, we focused on speaker conversion for transforming the voice identity of a source speaker into that of a target speaker as the most basic VC task, while gradually making their tasks more challenging, e.g., from parallel training (i.e., supervised training) in VCC 2016 into nonparallel training (i.e., unsupervised training) in VCC 2018. As described in Section 2, VC techniques were significantly improved through the activities of these challenges.

In 2020, we organized the third edition of the VCC, VCC 2020. As a more challenging task than the previous ones, we focused on cross-lingual VC, in which the speaker identity is transformed between two speakers uttering different languages, which requires handling completely nonparallel training over different languages. More details on the VCC 2020 task settings are given in Section 3. During the challenge period, more than 30 systems were developed with various VC techniques, such as neural vocoders, encoder-decoder networks, generative adversarial networks (GANs), and sequence-to-sequence (seq2seq) mapping networks, as summarized in Section 4. As in the previous challenges, we conducted large-scale listening tests to perceptually evaluate the voices converted by the individual systems. We observed from the results that further technical improvements were achieved in this challenge compared with in the previous ones as shown in Section 5. We also analyzed these results to investigate the effects of language differences on VC performance evaluation as presented in Section 6.

\section{Past voice conversion challenges and what we learned}

Before we introduce a new database and two new tasks for VCC 2020, we overview the past VCCs in 2016 and 2018 and what we learned. 

\subsection{Overview of the 2016 Voice Conversion Challenge}

The 2016 edition \cite{toda2016voice} was held as a special session of Interspeech 2016. We constructed a parallel VC database\footnote{The VCC2016 dataset is available for free at \url{https://doi.org/10.7488/ds/1575}}. Seventeen participants constructed their conversion systems by using the database. The parallel dataset consisted of four native speakers of American English (two females and two males), and each speaker uttered 162 common sentences. We used two of them as target speakers and the other two as source speakers. The participants were asked to produce converted speech for all possible source-target pairs. The number of converted audio samples per speaker was 54. 

The evaluation methodology was standard subjective evaluation based on listening tests. We evaluated the naturalness and speaker similarity of the samples converted to target speakers. For naturalness, we used the standard five-point-scale mean-opinion score (MOS) test ranging from 1 (completely unnatural) to 5 (completely natural). For the speaker similarity test, we adopted the same/different paradigm described in \cite{wester2016analysis}. Subjects were asked to listen to two audio pairs and to judge if they were the same speaker or not on a four-point scale: ``Same, absolutely sure,'' ``Same, not sure,'' ``Different, not sure,'' and ``Different, absolutely sure.'' More details on VCC 2016 can be found at \cite{wester2016analysis}. It was reported that the best system at that time obtained an average of 3.0 in the five-point-scale evaluation for the judgement of naturalness, and about 70\% of its converted speech samples were judged by listeners to be the same as the target speakers. However, it was obvious that there was a huge gap between the target natural speech and any of the converted speech. 

\subsection{Overview of the 2018 Voice Conversion Challenge}

The 2018 edition \cite{Lorenzo-Trueba2018} was held as a special session of ISCA Speaker Odyssey Workshop 2018. We constructed a new but smaller parallel VC database and a non-parallel VC database\footnote{The VCC2018 dataset is available for free at \url{https://doi.org/10.7488/ds/2337}.}. Twenty-three participants constructed their conversion systems by using the databases. The target and source speakers were four native speakers of American English, two females and two males, respectively, but they were different speakers from those used for the 2016 challenge. Each speaker uttered 80 sentences. Like the 2016 challenge, the participants were asked to produce and submit converted data for all possible source-target pairs. The number of test sentences for evaluation was 35. The same evaluation methodology as the 2016 challenge was adopted for the 2018 challenge, and more details can be found in \cite{Lorenzo-Trueba2018}. 

In the 2018 edition, we observed significant progress compared with the VCC 2016 results, and it was reported that, in both the parallel and non-parallel conversion tasks, the best system, using a phone encoder and neural vocoder, obtained an average of 4.1 in the five-point-scale evaluation for the judgement of naturalness, and about 80\% of its converted speech samples were judged by listeners to be the same as the target speakers. The best performing systems had similar performance in both the parallel and non-parallel conversion tasks. However, it was confirmed that there were statistically significant differences between the target natural speech and the best converted speech in terms of both naturalness and speaker similarity. 

\section{Tasks, databases, and timeline for Voice Conversion Challenge 2020}

\begin{figure}[t]
	\centering
	\includegraphics[width=1.0\linewidth]{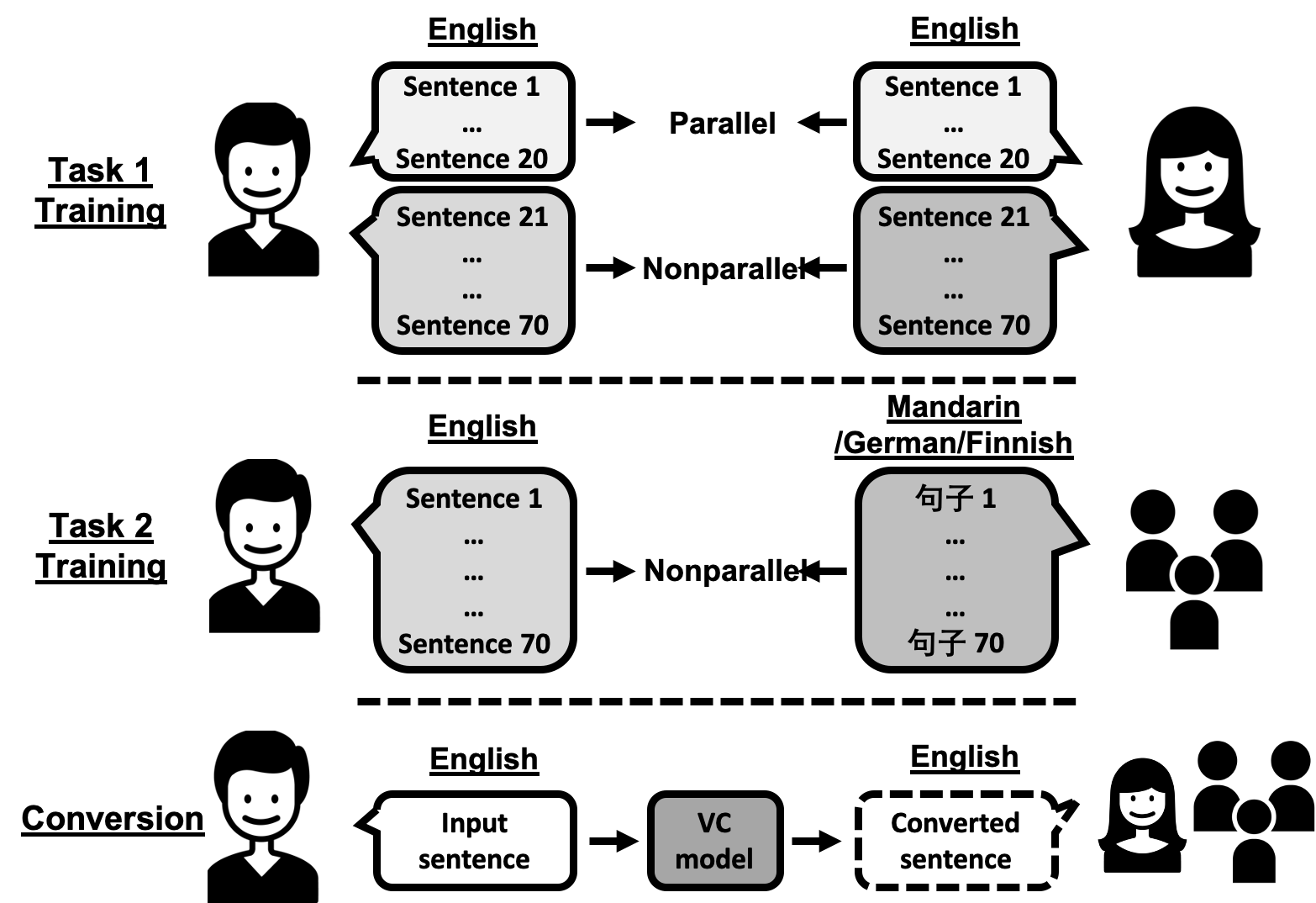} 
	\caption{Illustration of the two tasks in VCC 2020.	\label{fig:tasks}}
\end{figure}

The objective of VCC 2020, as in the past two challenges, is speaker conversion, which means that the converted speech should sound like the desired target speaker, with the same linguistic content as the source sentence. In this section, we explain two new tasks and the construction of the database in detail. Figure~\ref{fig:tasks} illustrates the two tasks.

\subsection{Task 1: Intra-lingual semi-parallel VC}

The dataset for Task 1 consists of a smaller parallel corpus (sentences uttered by both source and target speakers with the same content) and a larger nonparallel corpus, where both of them are of the same language. This setting is somehow realistic. Although it is impractical to get hundreds of parallel recordings, the use of a limited amount of parallel recordings may be acceptable. Although methods for nonparallel VC can be directly applied, it is expected that a small set of parallel sentences can aid in model learning. This task was introduced so we can assess the progress of VC systems compared with the past editions. 

\subsection{Task 2: Cross-lingual VC}

The dataset for Task 2 consists of a corpus of the source speakers speaking in the source language and another corpus of the target speakers speaking in the so-called \textit{target} language, although the aim is not to convert to this language. Since the two corpora are of different languages, the utterances are consequently different in content; thus, this task is nonparallel in nature. Using this database, participants in the challenge are supposed to disentangle speaker characteristics and the content of the source-speech data in the source language and to replace its speaker characteristics with those of the given target speaker regardless of what target languages the target speakers use. In fact, there are multiple target languages in Task 2, and they are Finnish, German, and Mandarin. This task is challenging in that the training dataset does not contain any \textit{source} language recording of the target speakers. In fact, such ground truth data can never be accessed.

\begin{table*}[t!]
\footnotesize
\centering
\caption{\label{table_aff} {List of participant affiliations of VCC 2020. They are listed in random order. }}
\begin{tabular}{|l|c|c|}
\hline
\multicolumn{1}{|c|}{\textbf{Affiliation}}                                                                      & \textbf{Task 1} & \textbf{Task 2} \\ \hline \hline
Academia Sinica                                                                                         & Y              & Y               \\ 
Beijing Forestry University & Y               & Y               \\ 
BUT \& UEF                                                                                                      & Y               & Y               \\ 
ChangHong                                                                                        & Y               & Y               \\ 
Chinese Academy of Science (Institute of Automation)                                                                                     & Y               & Y               \\ 
Chinese Academy of Sciences (Institute of Information Engineering)                                               & Y               & Y               \\ 
Duke Kunshan University                                                                                         & Y               & Y               \\ 
Dhirubhai Ambani Institute of Information and Communication Technology (DA-IICT)                                & Y               & Y               \\ 
Federal University of Rio de Janeiro                                                                            & Y               & N               \\ 
Guangdong University of Technology                                                                              & Y               & Y               \\ 
iQIYI Inc.                                                                                                      & Y               & Y               \\ 
Japan Advanced Institute of Science and Technology                                                              & Y               & Y               \\ 
KAKAO Enterprise                                                                                                & Y               & Y               \\ 
Logitech                                                                                           & N               & Y               \\ 
Microsoft                                                                                                       & Y               & Y               \\ 
MINDsLab Inc.                                                                                                   & Y               & N               \\ 
National Institute of Informatics                                                                               & Y               & Y               \\ 
National University of Singapore \& Northwestern Polytechnical University                                       & Y               & Y               \\ 
Nagoya University                                                                                               & Y               & Y               \\ 
NCSOFT                                                                      & N               & Y               \\ 
NetEase                                                                                            & Y               & N               \\ 
Nefrock                                                                                                         & Y               & N               \\ 
Peking University                                                                                                    & Y               & Y               \\ 
Personal (SSLab\_SRCB)                                                                                          & Y               & Y               \\ 
Personal (uVCTeam)                                                                                              & Y               & Y               \\ 
Red Pill Lab                                                                                                    & Y               & Y               \\ 
Spokestack.io                                                                                                   & Y               & N               \\
Tsinghua University                                                                                             & Y               & Y               \\ 
The University of Tokyo                                                                                         & Y               & Y               \\ 
University of Science and Technology of China                                                                   & Y               & Y               \\ \hline
\end{tabular}
\vspace{2mm}
\\ 
\end{table*}

\subsection{Database construction}

The VCC 2020 database is based on the Effective Multilingual Interaction in Mobile Environments (EMIME) dataset \cite{emime}, which is a bilingual database of Finnish/English, German/English, and Mandarin/English data. There are seven male and seven female speakers for each language, English, Finnish, German, and Mandarin, ending up in 56 speakers in total. The 145 sentences of the bilingual speakers were recorded in a semi-anechoic chamber. The recordings were down-sampled to 24 kHz and segmented into individual sentences.

As shown in Figure~\ref{fig:tasks}, the source languages in both tasks were set to English, and we chose two male and two female English speakers as the source speakers. The target language was also set to English in Task 1, while in Task 2, it was set to Finnish, German, and Mandarin as described above. From the remaining English speakers, we then chose another two male and two female speakers as the target speakers for Task 1. The criterion for choosing the English speakers was to make the speakers as perceptually discriminative as possible so that subjective conversion similarity would be easier to assess. In particular, we considered factors such as speaking rate, accent and overall pitch level. Then, we chose one male and one female for each of Finnish, German, and Mandarin. The criterion here was their fluency in English. We chose speakers who had the highest fluency scores according to an early study using this database \cite{emime}. 

Each of the source and target speakers has a training set of 70 sentences, which is around 5 minutes of speech data. Note that, in Task 1, the target and source speakers have 20 parallel sentences, where the remaining 50 sentences are different. The test sentences for evaluation are shared for Tasks 1 and 2 with a number of 25, and the sentences were released to participants about one week before they were required to submit their converted voices. The participants were asked to build systems for the $4\times 4=16$ and $4\times 6=24$ source-target pairs in Tasks 1 and 2, respectively.

 \subsection{Timeline}
 
The training databases for both tasks were released on March 9$^{\text {th}}$, 2020. The participants were given two months and two weeks to train their VC systems. Evaluation data was released on May 22$^{\text {nd}}$, 2020, and they were asked to upload their converted audio by May 29$^{\text {th}}$, 2020. They were also asked to submit system descriptions giving details on their constructed systems.  

\begin{table*}
\caption{\label{table_summary} Details of systems built by participants for VCC 2020. T11, T16, and T22 are baseline systems built by organizers. T18 and T30 breached rules, so information for their systems is not provided. }
\centering
\footnotesize
\begin{tabular}{|c|l|l||l|l|} 
\hline
\multirow{2}{*}{\textbf{Team ID }} & \multicolumn{2}{c||}{\textbf{Task 1}}                                            & \multicolumn{2}{c|}{\textbf{~Task 2} }                                            \\ 
\cline{2-5}
                                  & \multicolumn{1}{c|}{\textbf{VC model~}} & \multicolumn{1}{c||}{\textbf{Vocoder}} & \multicolumn{1}{c|}{\textbf{VC model~}} & \multicolumn{1}{c|}{~\textbf{Vocoder}}  \\ 
\hline 
T01                                & PPG-VC (Tacotron)~                      & Parallel WaveGAN                      & N/A                                     & N/A                                     \\
T02                                & PPG-VC (Tacotron)                       & WaveGlow                              & PPG-VC (Tacotron)                       & WaveGlow                                \\
T03                                & AutoVC                                  & WaveRNN                               & AutoVC                                  & WaveRNN                                 \\
T04                                & VQVAE                                   & WaveNet                               & N/A                                     & N/A                                     \\
T05                                & N/A                                     & N/A                                   & PPG-VC (IAF)                            & WORLD \& WaveGlow                         \\
T06                                & StarGAN                                 & WORLD                                 & StarGAN                                 & WORLD                                   \\
T07                                & NAUTILUS (Jointly trained TTS VC)        & WaveNet                               & NAUTILUS (Jointly trained TTS VC)        & WaveNet                                 \\
T08                                & VTLN + Spectral differential            & WORLD                                 & VTLN + Spectral differential            & WORLD                                   \\
T09                                & AutoVC                                  & Parallel WaveGAN                      & AutoVC                                  & Parallel WaveGAN                        \\
T10                                &   ASR-TTS (Transformer) / PPG-VC (LSTM)   & WaveNet                               & PPG-VC (LSTM)                           & WaveNet                                 \\
T11                                & PPG-VC (LSTM)                           & WaveNet                               & PPG-VC (LSTM)                           & WaveNet                                 \\
T12                                & ADAGAN                                  & AHOcoder                              & ADAGAN                                  & AHOcoder                                \\
T13                                & PPG-VC (Tacotron)                       & WaveNet                               & PPG-VC (Tacotron)                       & WaveNet                                 \\
T14                                & One shot VC                             & NSF                                   & N/A                                     & N/A                                     \\
T15                                & N/A                                     & N/A                                   & AutoVC                                  & MelGAN                                  \\
T16                                & CycleVAE                                & Parallel WaveGAN                      & CycleVAE                                & Parallel WaveGAN                        \\
T17                                & Cotatron                                & MelGAN                                & N/A                                     & N/A                                     \\
T19                                & VQVAE                                   & Parallel WaveGAN                      & VQVAE                                   & Parallel WaveGAN                        \\
T20                                & VQVAE                                   & Parallel WaveGAN                      & VQVAE                                   & Parallel WaveGAN                        \\
T21                                & CycleGAN                                & MelGAN                                & N/A                                     & N/A                                     \\
T22                                & ASR-TTS (Transformer)                   & Parallel WaveGAN                      & ASR-TTS (Transformer)                   & Parallel WaveGAN                        \\
T23                                & Transformer VC (Jointly trained TTS VC)  & Parallel WaveGAN                      & CycleVAE                                & WaveNet                                 \\
T24                                & PPG-VC (Tacotron)                       & LPCNet                                & PPG-VC (Tacotron)                       & LPCNet                                  \\
T25                                & PPG-VC (CBHG)                           & WaveRNN                               & PPG-VC (CBHG)                           & WaveRNN                                 \\
T26                                & One shot VC                             & Griffin-Lim                           & One shot VC                             & Griffin-Lim                             \\
T27                                & ASR-TTS (Transformer)                   & Parallel WaveGAN                      & PPG-VC / ASR-TTS (Transformer)          & Parallel WaveGAN                        \\
T28                                & Tacotron                                & WaveRNN                               & Tacotron                                & WaveRNN                                 \\
T29   & PPG-VC (CBHG)                           & LPCNet                               & PPG-VC (CBHG)                           & LPCNet                                 \\
T31                                & Multi-speaker Parrotron                 & WaveGlow                              & Multi-speaker Parrotron                 & WaveGlow                                \\
T32                                & ASR-TTS (Tacotron)                       & WaveRNN                               & ASR-TTS (Tacotron)                       & WaveRNN                                 \\
T33                                & ASR-TTS (Tacotron)                      & Parallel WaveGAN                      & PPG-VC (Transformer)                    & Parallel WaveGAN                        \\
\hline
\end{tabular}
\end{table*}

\section{Participants and submitted systems}
\label{participants_systems}
In this section, we briefly present the participants from both academy and industry (Section~\ref{participants}), followed by a summary of the submitted systems regarding both feature conversion (Section~\ref{feature_conversion}) and waveform generation methods (Section~\ref{vocoder}).

\subsection{Challenge participants}
\label{participants}

Table~\ref{table_aff} shows a list of participant affiliations and in which tasks they participated. They are listed in random order. In total, we received 33 submissions, including 3 baselines, from participants. Specifically, 31 teams submitted their results to Task 1, and 28 teams submitted their results to Task 2. There were 26 teams that participated in both tasks. In Sections~\ref{feature_conversion} and \ref{vocoder}, we briefly introduce the feature conversion and waveform generation methods of the submitted systems. 
Note that there were two teams who unfortunately did not submit the appropriate system descriptions despite repeated warnings from the organizers. Therefore, we exclude them in the following sections. 

Since the aim of the VCC is the scientific analysis of different VC methods, we assigned anonymized Team IDs (T01 to T33) to them. They are different from alphabetic order and also from the order of Table \ref{table_aff}. 

\subsection{VC systems built by challenge participants and baseline systems}
\label{baseline}

VC systems typically contain two modules, feature conversion and waveform generation. Both of them mainly use neural networks and there are several major approaches \cite{sisman2020overview}. 

Table \ref{table_summary} shows details on the two components used by participants for VCC 2020. The converted audio samples of all of the systems are available\footnote{Audio samples of baseline T11 are available at \url{https://www.dropbox.com/s/ansri259qqe3tpk/VCC20_refs.zip}. The converted samples of baseline T16 and T22 are available at \url{https://drive.google.com/drive/folders/1tboF-XCDlrxrB6CxUt9rojXRwc5K-LVv} and \url{https://drive.google.com/drive/folders/1C2BlumRiSNPsOCHgJNZVhpCbOXlBTT1w}}. Among the systems, T11, T16, and T22 are baseline systems built by the organizers. T11 uses the same configuration as the best performing system in VCC 2018. T16 is the CycleVAE + Parallel WaveGAN system \cite{tobing2019non}, which combines a representative encoder-decoder network-based method with a neural vocoder. T22 is a simple cascade of state-of-the-art ASR and TTS systems using seq2seq models, which have shown promising results in the past. Their converted audio were shared with all challenge participants. The latter two systems were implemented using open-source toolkits\footnote{Baseline T16 is based on open-source code available at \url{https://github.com/bigpon/vcc20_baseline_cyclevae}. Baseline T22 was implemented using the end-to-end speech processing toolkit ``ESPNet'' \cite{9053512} \url{https://github.com/espnet/espnet/tree/master/egs/vcc20}.}.  

Since the table includes many instances of jargon, it is not straightforward to derive any meaningful tendencies or scientific differences. We therefore grouped each of the two components into several coarse subgroups.

\subsection{Feature Conversion}
\label{feature_conversion}

Table~\ref{table_model} presents the models for feature conversion used in the participants' systems. According to the submitted system descriptions, the feature conversion models can be grouped into three sub-categories: 1) encoder-decoder model, 2) generative adversarial network (GAN) based model, and 3) parallel spectral feature mapping model. In general, the encoder-decoder and GAN-based models can be utilized with non-parallel data, while source-target paired data is required for the parallel spectral feature mapping model.

It is also noted that most teams used the same conversion model for both Task 1 and Task 2, while some teams, e.g.,\ T23 and T33, chose different solutions. It was also observed that teams 10 and 27 used two conversion models for a single task. In the following subsections, we summarize the three groups. 

\begin{table*}[t!]
\centering
\caption{\label{table_model} {Summary of feature conversion models used in submitted systems for Task 1 and Task 2. } 
}
\footnotesize
\begin{tabular}{|c|l|l|l|}
\hline
\textbf{Category}      &                                                                       \multicolumn{1}{c|}{\textbf{Conversion model}} & \multicolumn{1}{c|}{\textbf{Team ID (Task 1)}}                                                 & \multicolumn{1}{c|}{~\textbf{Team ID (Task 2)}}                                                                  \\ \hline \hline

                                                                        & PPG-VC                    & \begin{tabular}[c]{@{}l@{}} T02, T10, T11, T13, T24, \\ T25, T29, T31 \end{tabular} & 
                                                                    \begin{tabular}[c]{@{}l@{}} T02, T05, T10, T11, T13, T24, \\ T25, T27, T29, T31, T33 \end{tabular}                                                                                \\ \cline{2-4}
                                                                        
                                      Encoder-decoder                                 & ASR-TTS                    & T10, T22, T27, T32, T33                                                            & T22, T27, T32                                                                              \\ \cline{2-4} 

                        (non-parallel data)
                       & Leverage TTS for VC   & T07, T23, T17                                                                      & T07                                                                                        \\ \cline{2-4} 
                                                                                               & AutoEncoder                & \begin{tabular}[c]{@{}l@{}}T03, T04, T09, T14, T16, T19, \\ T20 , T26\end{tabular} & \begin{tabular}[c]{@{}l@{}}T03, T09, T15, T16, T19, T20, \\ T23, T26\end{tabular}                    \\ \hline
\multirow{1}{*} GAN-based Model & CycleGAN                   & T21                                                                                & N/A                                                                                        \\ \cline{2-4} 
                                                                       (non-parallel data)                        & StarGAN                    & T06                                                                                & T06                                                                                        \\ \cline{2-4} 
                                                                                               & ADAGAN                     & T12                                                                                & T12                                                                                        \\ \hline
\multirow{1}{*} Parallel Spectral Mapping     & Tacotron                   & T01, T28                                                                           & T28                                                                                        \\ \cline{2-4} 
                                                                       (parallel data)                        & VTLN+spectral differential & T08                                                                                & T08                                                                                        \\ \hline
\end{tabular}
\end{table*}

\subsubsection{Encoder-decoder model}

In the encoder-decoder model sub-category, a speech signal is first encoded into speaker independent (SI) features representations, e.g.,\ Phonetic PosteriorGram (PPG)~\cite{sun2016phonetic, tian2018average, Liu2018} and text or other latent content code~\cite{qian2019autovc, van2017neural, chou2019one, hsu2017voice}. Then, the decoder is to predict the corresponding acoustic features or time-domain speech signals with the SI features. At run-time, the same SI features extracted by the encoder from a given speech input are used to drive the decoder to generate the converted feature or speech signal. As the decoder is trained to perform a mapping between the SI features and the corresponding acoustic features or speech signals of the same speaker, parallel data is not needed for model training. The encoder and decoder are generally trained with a multi-speaker corpus, and we can control the speaker identity of synthetic speech by either fine-tuning the decoder model with a small amount of target speech or by conditioning the decoder on the basis of a speaker identity vector, e.g.,\ one-hot vector, i-vector~\cite{dehak2010front}, x-vector~\cite{snyder2018x}, and similar neural speaker embedding.

\begin{table*}[t]
\centering
\caption{\label{table_vocoder} {Summary of vocoders used in submitted systems for Task 1 and Task 2. }}
\footnotesize
\begin{tabular}{|c|l|l|l|} 
\hline
\textbf{Type}                                                                                            & \multicolumn{1}{c|}{\textbf{Vocoder}} & \multicolumn{1}{c|}{\textbf{Team ID (Task 1)}}                                                 & \multicolumn{1}{c|}{~\textbf{Team ID (Task 2)}}                                            \\ 
\hline \hline
\multirow{1}{*} Neural Vocoder      & WaveNet               & T04, T07, T10, T11, T13   & T07, T10, T11, T13, T23  \\ 
\cline{2-4}
            (Autoregressive)        & WaveRNN               & T03, T25, T28, T32        & T03, T25, T28, T32       \\ 
\cline{2-4}
                                    & LPCNet                & T24, T29                  & T24, T29                 \\ 
\hline
\multirow{2}{*}                     & Parallel WaveGAN      & \begin{tabular}[c]{@{}l@{}}T01, T09, T16, T19, T20, \\
                                                            T22, T23, T27, T33\end{tabular} 
                                                            & \begin{tabular}[c]{@{}l@{}} T09, T16, T19, T20, T22, \\
                                                            T27, T33\end{tabular}  \\ 
\cline{2-4}
                Neural Vocoder      & WaveGlow              & T02, T31                  & T02, T05 (denoising), T31 \\ 
\cline{2-4}
          (Non-autoregressive)      & MelGAN                & T17, T21                  & T15  \\ 
\cline{2-4}
                                    & NSF                   & T14                       & N/A   \\ 
\hline
\multicolumn{1}{|l|}{\multirow{1}{*}}  
                                    & WORLD                 & T06, T08                  & T05, T06, T08 \\ 
\cline{2-4}
\multirow{1}{*} Traditional Vocoder & AHOcoder              & T12                       & T12 \\ 
\cline{2-4}
\multicolumn{1}{|l|}{}              & Griffin-Lim           & T26                       & T26  \\
\hline
\end{tabular}
\end{table*}

According to Table~\ref{table_model}, the encoder-decoder model structure was the most popular among the submitted systems, where 23 and 22 teams reported having based their work on this structure for the monolingual and cross-lingual conversion tasks, respectively. The encoder-decoder model structure used by the participants can be further divided into four types:

\leftmargini=5mm
\begin{itemize}
    \item[1)] \textbf{PPG-VC:} In this framework, speech is first encoded into a frame-level phonetic information representation. A decoder is then trained to transform the PPGs to target speech. For most systems, the PPG is trained with monolingual speech~\cite{tian2018average, Liu2018}, while, for system T13, bilingual PPG~\cite{zhou2019cross} is used to capture the phonetic information of different languages. For the decoder, various network structures are reported, e.g.,\ Long Short Term Memory network (LSTM)~\cite{Liu2018}, Tacotron~\cite{wang2017tacotron}, Transformer~\cite{vaswani2017attention, li2019neural}, CBHG~\cite{wang2017tacotron}, and Inverse Autoregressive Flow (IAF)~\cite{kingma2016improved}. In practice, the encoder and decoder can be optimized either separately or jointly (system T31)~\cite{biadsy2019parrotron}.
    
    \item[2)] \textbf{ASR-TTS:} In this framework, a pretrained automatic speech recognition (ASR) model is first used to recognize the text information of a speech signal. A text-to-speech (TTS) model, e.g.,\ Tacotron~\cite{wang2017tacotron} and Transformer TTS~\cite{li2019neural}, is then used to synthesize the target speech with the text sequence. As the temporal information is discarded during speech recognition, the duration of converted speech is also generated by TTS, which is usually different from that of a source speaker.
    
    \item[3)] \textbf{Leverage TTS for VC:} In this framework, TTS systems were used to boost the performance of VC systems. System T17 (Cotatron)~\cite{park2020cotatron} directly uses a pretrained TTS encoder to extract linguistic features. Then, the decoder takes the linguistic features as the input for speech generation. As the encoder is guided by the text, the text transcription of source speech is required at run-time. Alternatively, system T23 (Voice Transformer Network)~\cite{huang2019voice} first initializes the decoder with pretrained TTS decoder parameters. Then, a two-stage training is applied to optimize the encoder and decoder. Different from these two methods, system T07 (NAUTILUS)~\cite{luong2020nautilus} utilizes a joint architecture, which accepts either text or speech as input and optimizes the architecture for both TTS and VC tasks. The core idea is to force the speech encoder to learn the same speaker-disentangled latent linguistic embedding as the text encoder. At run-time, by only feeding the source speech as input, the shared decoder generates the converted speech.

    \item[4)] \textbf{Auto-encoder:} Auto-encoder based VC aims to decouple input speech into a speaker-independent latent representation and a speaker latent representation; then; a decoder learns to reconstruct the original signal with the latent representations. For conversion, a new speaker latent vector extracted from a target speaker and speaker-independent latent representations extracted from a source speaker are used as the inputs of the decoder. According to the submitted systems, four types of models were used: AutoVC~\cite{qian2019autovc}, VQ-VAE~\cite{van2017neural}, CycleVAE~\cite{tobing2019non}, and one-shot VC~\cite{chou2019one}.
\end{itemize}

\subsubsection{GAN-based model}
GAN-based models jointly train a generator network with a discriminator, where an adversarial loss derived from the discriminator is used to encourage the generator outputs to be indistinguishable from real speech features. Thanks to so-called cycle consistency training \cite{zhu2017unpaired}, GAN-based models can also be trained without using parallel data. CycleGAN-VC~\cite{kaneko2018cyclegan} is used in system T21 for one-to-one mapping. It is designed to learn a spectral mapping $G$ of source to target and its inverse mapping $F$ jointly. The network is optimized with a cycle consistency loss combining an adversarial loss and an identity mapping loss. StarGAN-VC~\cite{kameoka2018stargan} and adaptive GAN-based VC~\cite{patel2019adagan} are also used for many-to-many conversion, where a speaker identity vector is used as an additional input of generators to control the generated speech identity.

\subsubsection{Parallel spectral feature mapping model}
This group of models relies on parallel utterance pairs of source and target speakers to build a spectral feature mapping. Two submitted systems (T01 and T28) were based on the Tacotron structure, where a speaker embedding is concatenated with encoded output to present the speaker identity. System T01 utilizes additional parallel data for conversion model training, while system T28 artificially generates parallel data using a pretrained TTS system. A two-step conversion method is used in system T08. During training, a vocal tract length normalization (VTLN)~\cite{sundermann2003vtln} based warping function and linear pitch feature conversion are learned. At run-time, the spectral feature is converted by VTLN, and an intermediate waveform is generated with converted F0 with original speech features. Then, differential spectral compensation~\cite{kobayashi2018statistical} is employed for converted speech generation. For Task 1, the conversion model is trained with the 20 parallel sentences, while, for Task 2, the INCA algorithm~\cite{erro2009inca} is used to obtain aligned source-target feature pairs for model training.

\subsection{Waveform generation}
\label{vocoder}

The vocoder is another module for VC that plays a crucial role in generating speech waveforms from converted speech features. It affects the quality of converted speech significantly. Table~\ref{table_vocoder} presents a summary of the vocoders used in the submitted systems. They can be grouped into three categories: the auto-regressive neural vocoder, non-autoregressive neural vocoder, and traditional vocoder. In general, the neural vocoder works in a data-driven manner, which requires a big amount of data for training, while traditional vocoders are signal processing approaches. Training data is not required for such models. We summarize the three groups in the following subsections.

\subsubsection{Neural vocoder (autoregressive)}

The autoregressive neural vocoder is a generative network that directly models the relationships among time-domain speech samples. It predicts the distribution of a current sample conditioned on the previous generated samples and auxiliary speech features, e.g., mel-spectrum. In practice, the network can be implemented using either a convolutional neural network (CNN) or recurrent neural network (RNN). As shown in Table~\ref{table_vocoder}, 11 participants chose the autoregressive neural vocoder for speech signal generation. Specifically, five teams used WaveNet~\cite{oord2016wavenet}, while four and two teams used WaveRNN~\cite{kalchbrenner2018efficient} and LPCNet~\cite{valin2019lpcnet} in their system implementation, respectively.

\subsubsection{Neural vocoder (non-autoregressive)}

The non-autoregressive neural vocoder is another type of time-domain waveform modeling approach. In practice, different models are used in implementation, e.g., flow-based, GAN-based, and neural source-filter models. Due to the parallel generation mechanism, these models work much faster than their auto-regressive counterparts.
According to the submissions, 14 and 11 teams chose non-autoregressive neural vocoders for Task 1 and Task 2, respectively. Among the submissions, Parallel WaveGAN~\cite{yamamoto2020parallel} was the most popular one, where nine and seven teams chose it to generate converted speech for Task 1 and Task 2, respectively, while WaveGlow~\cite{prenger2019waveglow}, MelGAN~\cite{kumar2019melgan}, and the neural source-filter waveform model (NSF)~\cite{wang2019neural} were also used in some implementations. 

\subsubsection{Traditional vocoder}

Traditional vocoders are pure signal-processing methods, which are generally built on the basis of different assumptions to model the process of human speech generation. For example, the source-filter model is based on the assumption that a target spectrum can be obtained by modulating an excitation signal using a filter corresponding to the vocal tract, while the harmonic plus noise model (HNM) assumes that speech signals can be decomposed into a harmonic band and a noise-like band. Alternatively, Griffin-Lim~\cite{griffin1984signal} generates speech by reconstructing the phase on the basis of the spectrum magnitude. As shown in Table~\ref{table_vocoder}, there are two and three teams that used WORLD~\cite{morise2016world} for Task 1 and Task 2, respectively. Another two teams used AHOcoder~\cite{erro2013harmonics} and Griffin-Lim in their systems. 
  
We note that additional algorithms are also used for waveform generation. Specifically, for system T05, a WORLD vocoder is used to reconstruct speech signals from converted speech features, followed by a WaveGlow for denoising. System T13 also applies a vowel-focused denoising (WebRtc\_NS\footnote{\url{https://github.com/cpuimage/WebRTC_NS}}) after the WaveNet vocoder. Additionally, system T08 first uses the WORLD vocoder to generate an intermediate waveform; then, a time-domain spectral differential is employed for generating converted speech.

\subsection{Brief descriptions of T10}

Among those systems, team T10 obtained impressive results as we will describe in the next section. Fortunately, team T10 kindly provided us with a more detailed description of their system to make our analysis more meaningful.

For Task 1, the T10 system was constructed on the basis of two approaches. The first was ASR-TTS, which concatenated an ASR module with a TTS module. At the conversion stage, source speech was first fed into an English ASR engine provided by iFlytek to obtain transcriptions. Then, speaker-dependent transformer-based TTS models predicted mel-spectrograms from input transcriptions. Finally, speaker-dependent WaveNet vocoders, which used single Gaussian output distributions, reconstructed 24-kHz/16-bit waveforms from the predicted mel-spectrograms. Furthermore, a prosody encoder was connected with the transformer-based TTS model to extract sentence-level prosody code from the source speech and to use it at the conversion stage. Preliminary experiments showed that this modification can help to slightly improve the naturalness of converted speech for the conversion pairs in which TEF2 was the target speaker. Therefore, the prosody encoder was only utilized for these pairs in the submitted results for T10.

The second approach followed the PPG-VC framework \cite{sun2016phonetic} and was developed on the basis of the N10 system in VCC2018 \cite{Liu2018}. Several improvements were made. First, an autoregressive structure was introduced into LSTM-based acoustic feature predictors for generating mel-spectrograms from bottleneck features extracted by an English ASR acoustic model. Second, at the conversion stage, the sequences of the bottleneck features were interpolated to compensate for the mismatch between the speaking rates of two speakers. Third, 24-kHz/16-bit waveforms, instead of 16-kHz/10-bit ones, were recovered from the predicted mel-spectrograms by speaker-dependent WaveNet vocoders. 

In both approaches, the transformer-based TTS models, acoustic feature predictors, and WaveNet vocoders were pre-trained by a large multi-speaker dataset and then fine-tuned on the target speaker. For generating the final results, the first approach was the default one, and the second approach was adopted only by the conversion pairs of SEM1-TEM1 and SEM1-TEM2 according to internal evaluations on the development data.

For Task 2, the T10 system was constructed in the same way as the second approach for Task 1. Preliminary experiments found that the F0 contours of the conversion pairs SEF1-TGM1, SEM1-TGM1, and SEM2-TGM1 were not satisfactory. Therefore, the logarithmic F0s of source speech were linearly converted and then used for these pairs.

\section{Subjective evaluation}

One of the evaluation methodologies adopted for VCC 2020 is subjective evaluation based on listening tests. Here, we describe the design and results of the tests. We also carried out an objective evaluation, and it will be reported in a separate paper \cite{vcc2020objective}.

\subsection{Motivations and evaluation methodology}

What we want to find out is the naturalness and speaker similarity of the converted samples in a similar way to the evaluation methodology used in previous VCCs. However, the evaluation methodology needed to be refined so that we could evaluate cross-lingual VC properly. In particular, we changed the naturalness and speaker similarity evaluation in Task 2: 
\begin{itemize}
\item In addition to natural speech in English, natural speech in either German, Finnish, and Mandarin was also rated by subjects. 
\item In addition to reference speech in English, reference speech in either German, Finnish, and Mandarin was also presented to subjects for judging speaker similarity across languages. 
\end{itemize}

\subsection{Experimental setup}
\begin{figure}[t]
	\centering
	\includegraphics[trim=6cm 0.5cm 1cm 2cm,clip,width=1.0\linewidth]{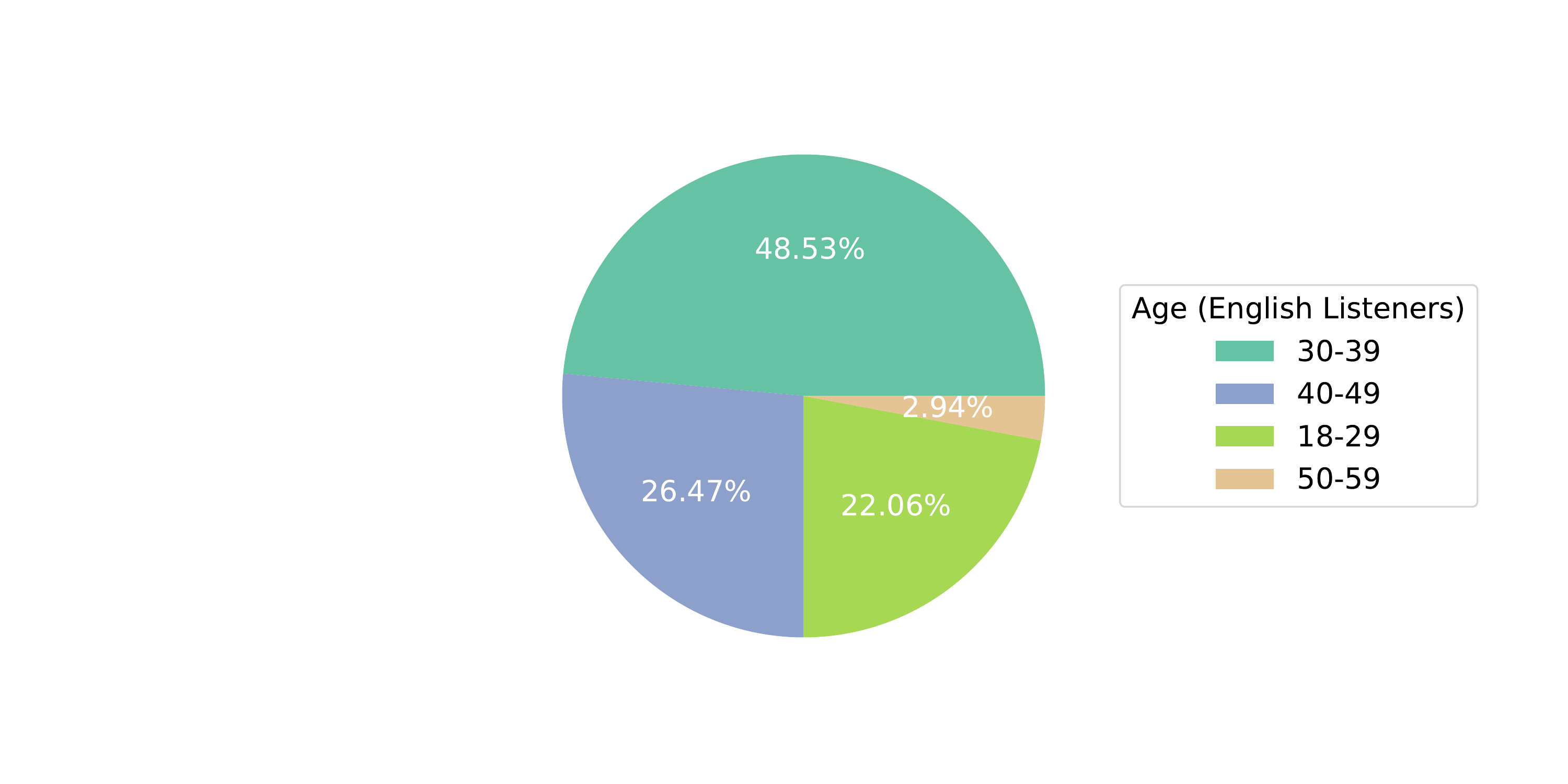} \\
	\includegraphics[trim=6cm 0.5cm 1cm 2cm,clip,width=1.0\linewidth]{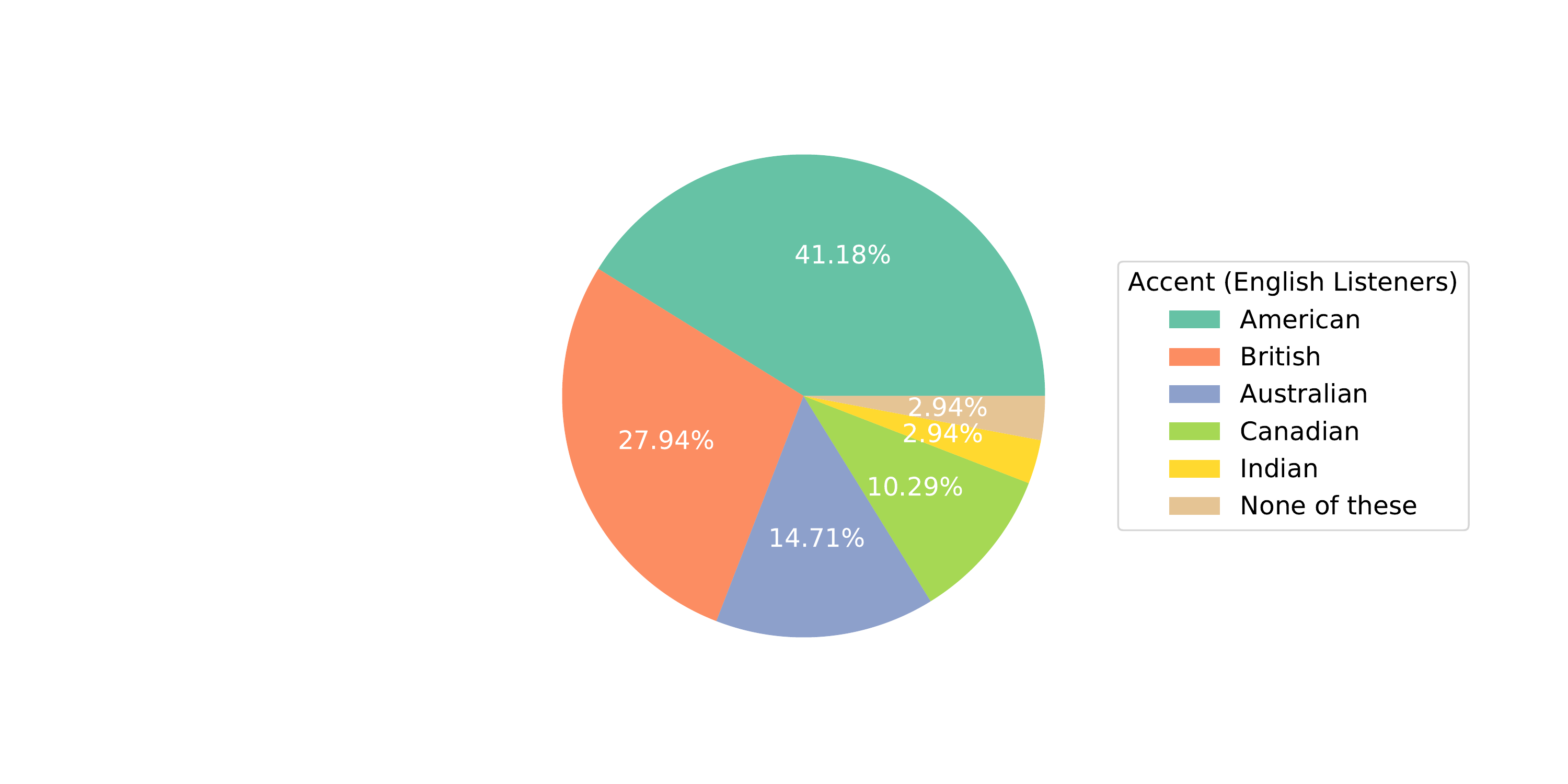} \\
	\includegraphics[trim=6cm 0.5cm 1cm 2cm,clip,width=1.0\linewidth]{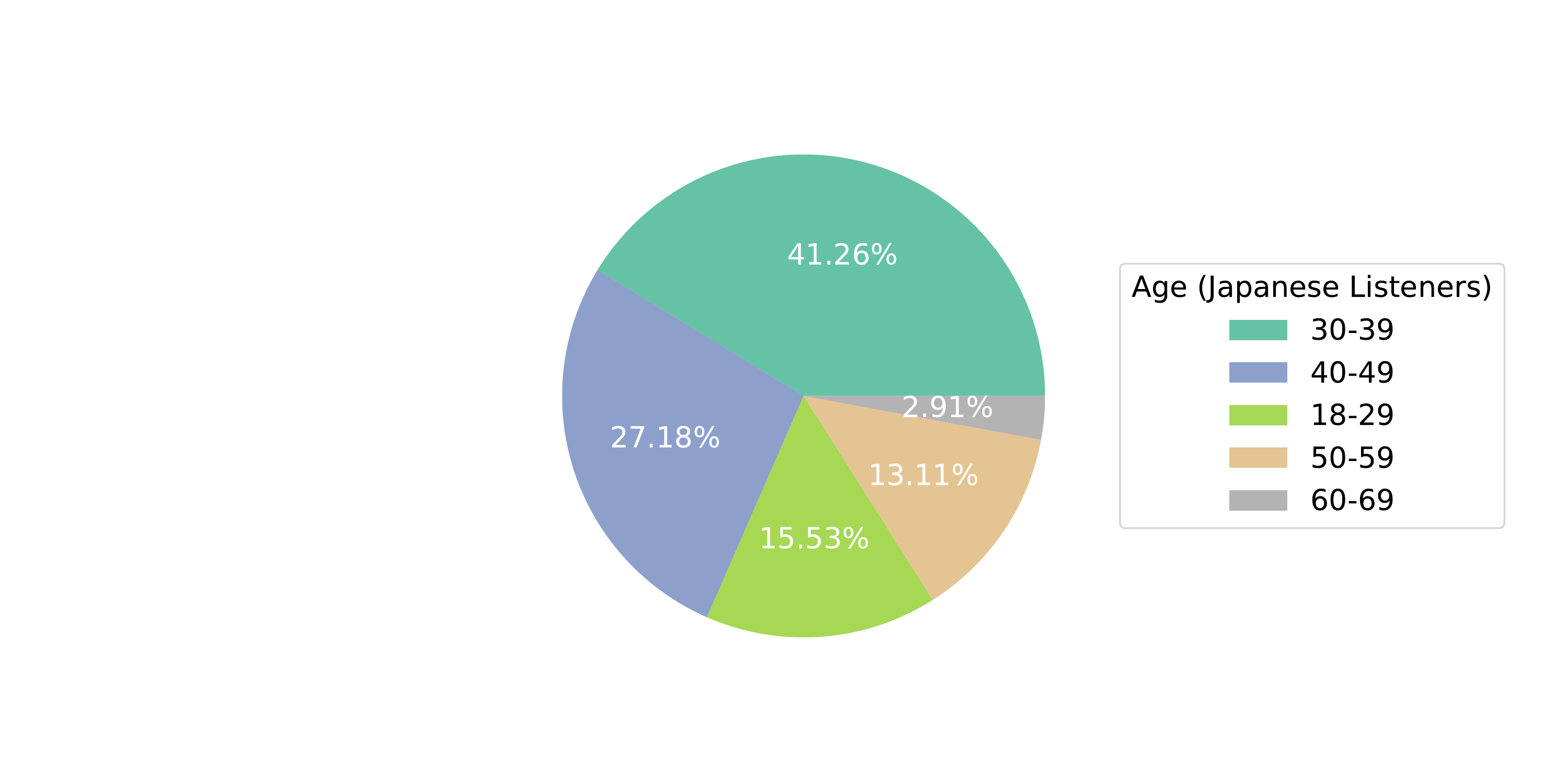} 
	\caption{Age and accent distribution of English and Japanese listeners.	\label{fig:en_accent}}	
\end{figure}

We also investigated different types of listeners. One of the main applications of cross-lingual VC is speech-to-speech translation. In such applications, speech-to-speech translation may be used for travelers (that is, German, Finnish, and Mandarin) to be able to communicate with people in foreign countries (e.g.,\ Japan), where English is used as or supposed to be a L2 language. In this scenario, the expected listeners may not be native speakers of English. Therefore, we recruited both native speakers of English and non-native speakers of English (in this case, Japanese) to observe the differences brought about by listeners.

\begin{figure*}[t]
	\centering
	\includegraphics[width=0.9\linewidth]{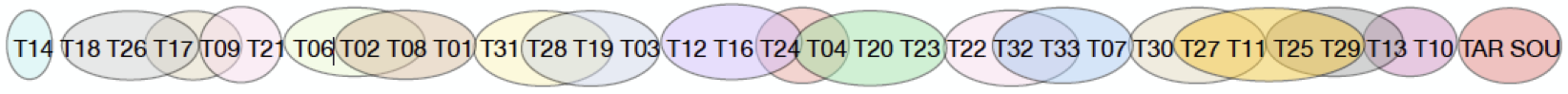} 
	\caption{Groupings of systems that did not differ significantly from each other in terms of naturalness for Task 1. }	\label{fig:en_intra_qua_sig}
\end{figure*}

\begin{figure*}[t]
	\centering
    \includegraphics[width=0.65\linewidth]{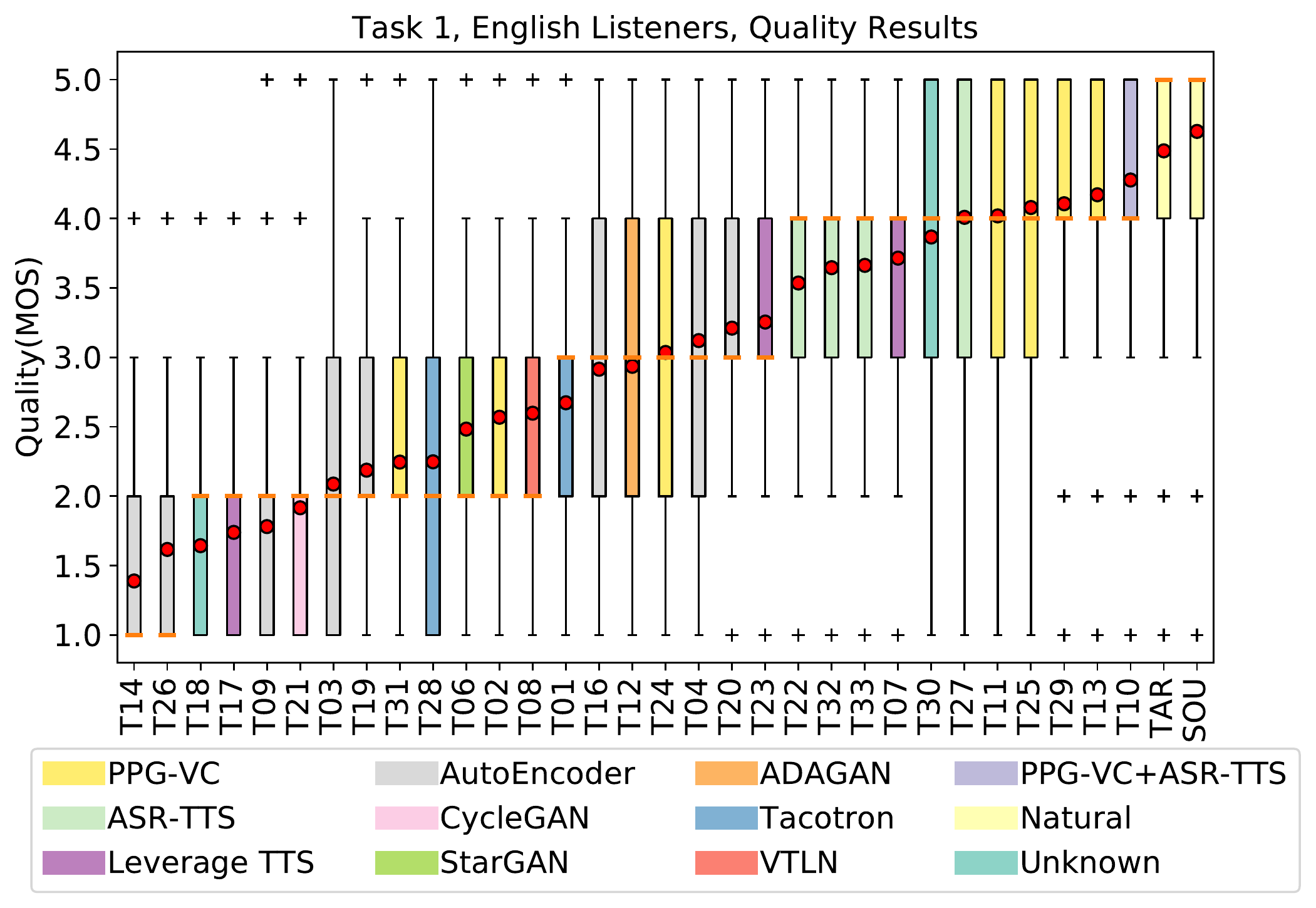} 
	\vspace{-2mm}
	\caption{Naturalness results for Task 1. MOS scores are arranged in accordance with their mean (red dot). Bars are colored on basis of feature conversion categories.}	\label{fig:en_intra_score_qua}
\end{figure*}

The evaluation instructions and scales given to the subjects were the same as for the previous VCCs. To evaluate naturalness, listeners were given the instruction below: 
\begin{quote}
Part 1: Listen to the following audio and rate it for quality. Some of the audio samples you will hear are of high quality, but some of them may sound artificial due to deterioration caused by computer processing. Please evaluate the voice quality on a scale of 1 to 5 from ``Excellent'' to ``Bad.'' Quality does not mean that the pronunciation is good or bad. If the pronunciation of the English is unnatural but the sound quality is very good, please choose ``Excellent.''
\end{quote}
They were then asked to rate how natural the speech sounded on a five-point scale: (1) Bad, (2) Poor, (3) Fair, (4) Good, and (5) Excellent. Again, in addition to natural speech in English, natural speech in either German, Finnish, or Mandarin was also rated by the subjects for Task 2.

To evaluate the speaker similarity of converted samples to reference audio, the same/different paradigm from VCC 2018 was used. The listeners were given the instruction below: 
\begin{quote}
Part 2: Please listen to the following two audio samples and rate them for speaker similarity. Please consider who is speaking according to the characteristics of the sound and then make a choice using a 4-level scale that varies from ``Same (sure)'' to ``Different (sure)'' to rate the speaker similarity of the two audio samples. Please do not consider the content or language to which you are listening.
\end{quote}
They were then asked to rate the speaker similarity of the two samples on a four-point scale: (4) same speaker, absolutely sure, (3) same speaker, not sure, (2) different speaker, not sure, (1) different speaker, absolutely sure. Again, the reference speech could be in English or either German, Finnish, or Mandarin for Task 2.

We subcontracted the crowd-sourced perceptual evaluation with English and Japanese listeners to Lionbridge Technologies Inc. and Koto Ltd., respectively. The two sets of perceptual evaluation required a total of \textyen{738,128} Japanese yen (approx.\ 6,900 USD). Please note that we did not ask the participants to pay this expensive fee\footnote{Challenge participants may be charged in future challenges.}.

\begin{figure*}[t]
	\centering
	\includegraphics[width=0.9\linewidth]{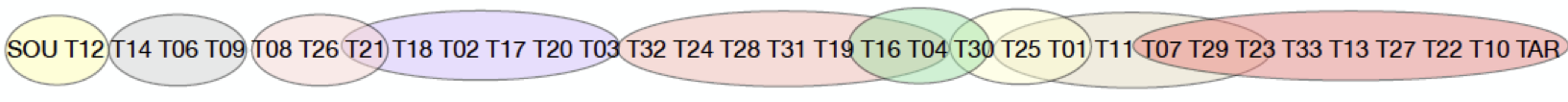} 
	\caption{Groupings of systems that did not differ significantly from each other in terms of similarity to target speaker for Task 1.  }	\label{fig:en_intra_sim_sig}
\end{figure*}

\begin{figure*}[t]
	\centering
	\includegraphics[trim=1cm 0 1cm 9,clip,width=0.7\linewidth]{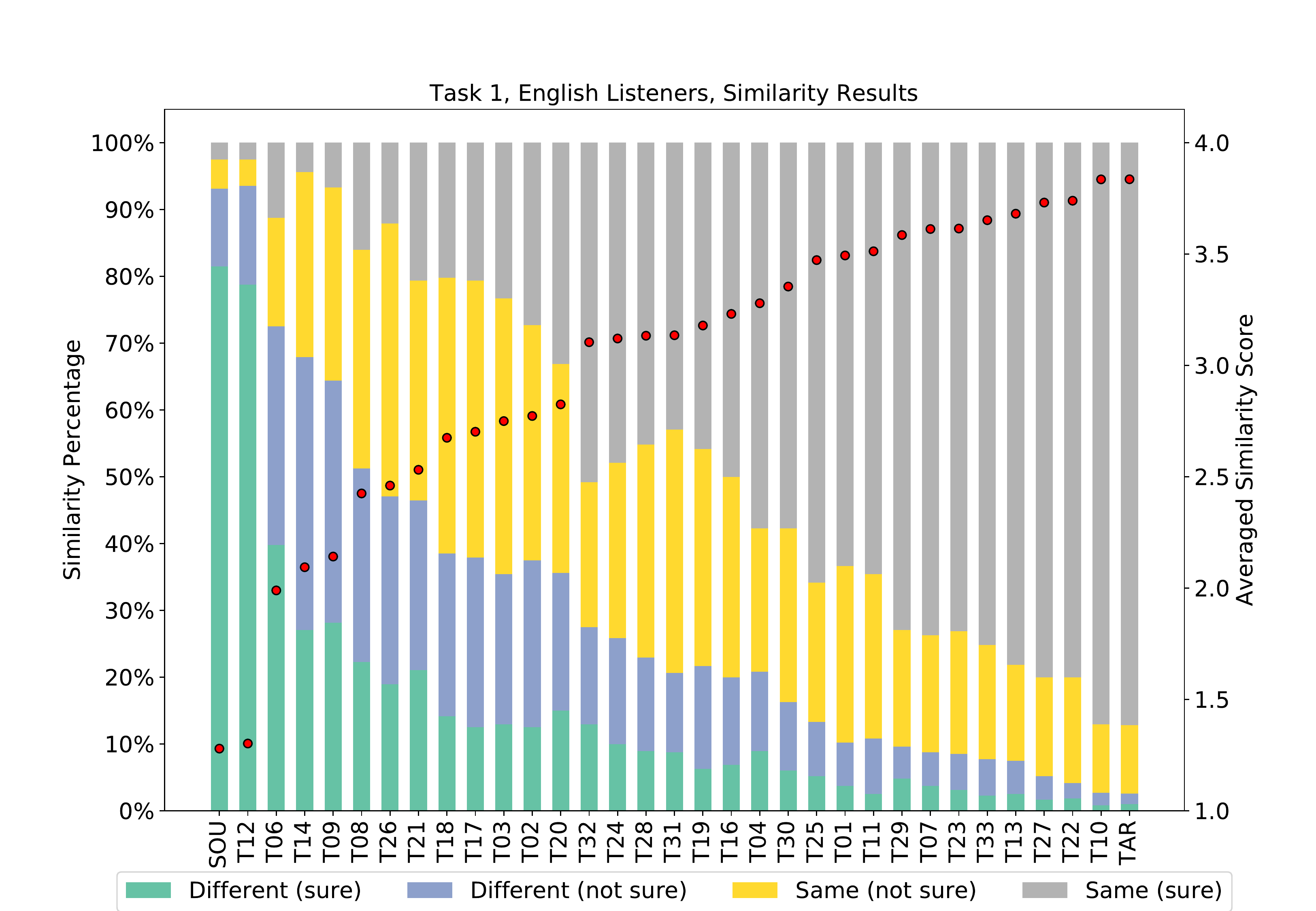} 
	\caption{Similarity results of the target speaker for Task 1. Similarity scores are arranged in accordance with their mean value (red dot).}\label{fig:en_intra_score_sim}
\end{figure*}

\begin{figure}[t]
	\centering
	\includegraphics[trim=1cm 1cm 1cm 1cm,clip,width=0.9\linewidth]{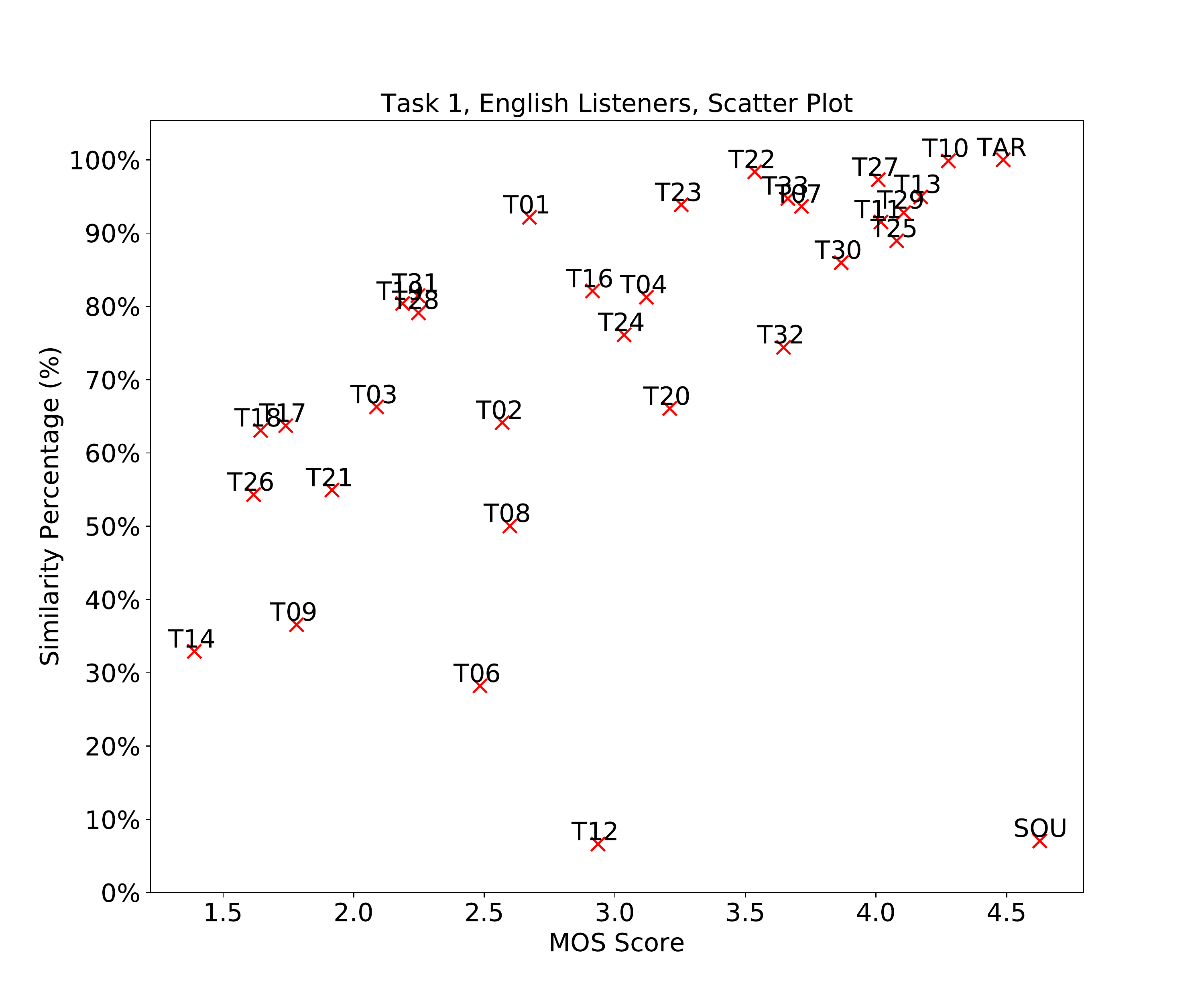} 
	\caption{Scatter plot matching naturalness and scores of similarity to target speaker for Task 1 when averaging all speaker pairs (y-axis is similarity percentage). \label{fig:en_intra_scatter}}
\end{figure}

Given the extremely large costs required for the perceptual evaluation, we selected 5 utterances (E30001, E30002, E30003, E30004, E30005) only from each speaker of each team. To evaluate the speaker similarity of the cross-lingual task, we used audio in both the English language and in the target speaker's L2 language as reference. For each source-target speaker pair, we selected three English recordings and two L2 language recordings as the natural reference for the converted five utterances.

Each evaluation set contained 62 webpages, and each webpage had 3 unique audio samples: one to evaluate its quality, one to evaluate its speaker similarity, and the other was used as a reference for evaluating speaker similarity. All of them corresponded to a total of 62 different systems: 31 for intra-lingual and 28 for cross-lingual systems; 1 system was for evaluating target samples, and 2 systems were used to evaluate source samples (one system was used for comparison with the intra-language reference and another for comparison with the cross-lingual reference).

Each audio sample in Task 1 was rated 6 times, and each sample in Task 2 was rated 4 times. Each source speech sample was evaluated at least 4 times, and each target sample was evaluated 12 times. In total, 480 evaluation sets were employed to cover all 29,760 data points. Before doing the experiment, all of the VC audio files were converted to 24 kHz and 16-bit precision in signed integer PCM format.

We prepared two identical experiments, and the two subcontracted companies recruited English and Japanese listeners, respectively. For the experiment participated in by English subjects, we had a total of 68 unique valid listeners (32 female, 33 male, and 3 unknown), and they evaluated 472 sets, with an average of 7 sets per participant. For the experiment participated in by Japan listeners, we had a total of 206 unique valid listeners (96 male and 110 female), and they evaluated 475 sets, with an average of 2.31 sets per participant. Figure~\ref{fig:en_accent} shows the accent and age distribution of the English and Japanese listeners. We can see that almost half of the English participants were in their 30s or 40s, and most of them had American or British accents. We can also see that most of the Japanese listeners were also in their 30s or 40s, which was similar to the distribution of English listeners. 

In the following subsections, we will first show the results for the English subjects. 

\subsection{Evaluation results -- English listeners --}
\label{main-result}
\subsubsection{Task 1 - Naturalness}

Figure~\ref{fig:en_intra_score_qua} shows box plots and mean scores for the results of the naturalness evaluation averaged across all speaker pairs, and Figure~\ref{fig:en_intra_qua_sig} shows Wilcoxon signed-rank tests with Bonferroni correction ($\alpha$ = 0.01) in which systems are grouped that did not differ significantly from each other in Task 1. TAR and SOU refer to the natural speech of the target speaker and the source speaker, respectively. Again, T11, T16, and T22 were the baseline systems described in Section \ref{baseline}. The colors of the boxplot represent the feature conversion categories described in Section \ref{feature_conversion}. Colorization based on the vocoder types described in Section \ref{vocoder} is shown in Appendix \ref{apped:vocoder}.

\vspace{1mm}
\noindent 
\textbf{Comparison with T11:} From Figure~\ref{fig:en_intra_score_qua}, we can first see that T10 and T13 obtained the highest MOS values among the VC systems, and their improvements compared with T11, which was the best performing system in VCC 2018, were statistically significant. Strictly speaking, T10 was better than any of the other VC systems apart from T13, while T13 was not better than T10, T29, and T25 but significantly better than any of the other systems. We can therefore say that VC performance has improved within the past two years in terms of naturalness. Both of them use PPG-based approaches. We can also see that T29, T25, T27, and T30 were also as good as T11 and were not significantly different from T11.

\vspace{1mm}
\noindent 
\textbf{Comparison with humans:} T10 and T13 achieved the highest scores, but their scores were still lower than those for the natural speech of the source and target speakers, and their differences were statistically significant. In other words, VC systems have advanced within the past two years, but \textit{none of them have achieved human-level naturalness yet}. In terms of audio naturalness, basic intra-lingual VC has not been completely solved.  

\vspace{1mm}
\noindent 
\textbf{PPG vs TTS:} We can see that most of the best performing systems used PPG (T10, T13, T25, T29) entirely or partially. We can also see that VC systems using a combination of ASR and TTS systems (T27, T33, T32, and T22) are located in the upper group. It seems that as long as we can obtain ASR systems suitable for target speakers, the combination of ASR and TTS is also a reasonable approach for VC. 

\begin{figure*}[t]
	\centering
	\includegraphics[width=0.9\linewidth]{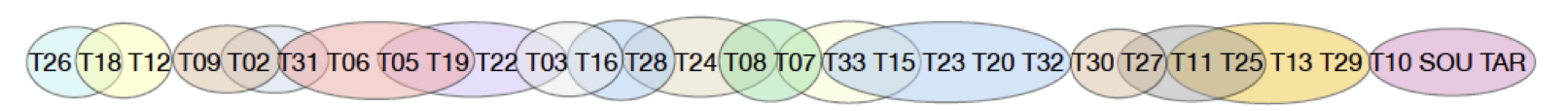} 
	\caption{ Groupings of systems that did not differ significantly from each other in terms of naturalness for Task 2. }	\label{fig:en_cross_qua_sig}
\end{figure*}
\begin{figure*}[t]
	\centering
	\includegraphics[width=0.65\linewidth]{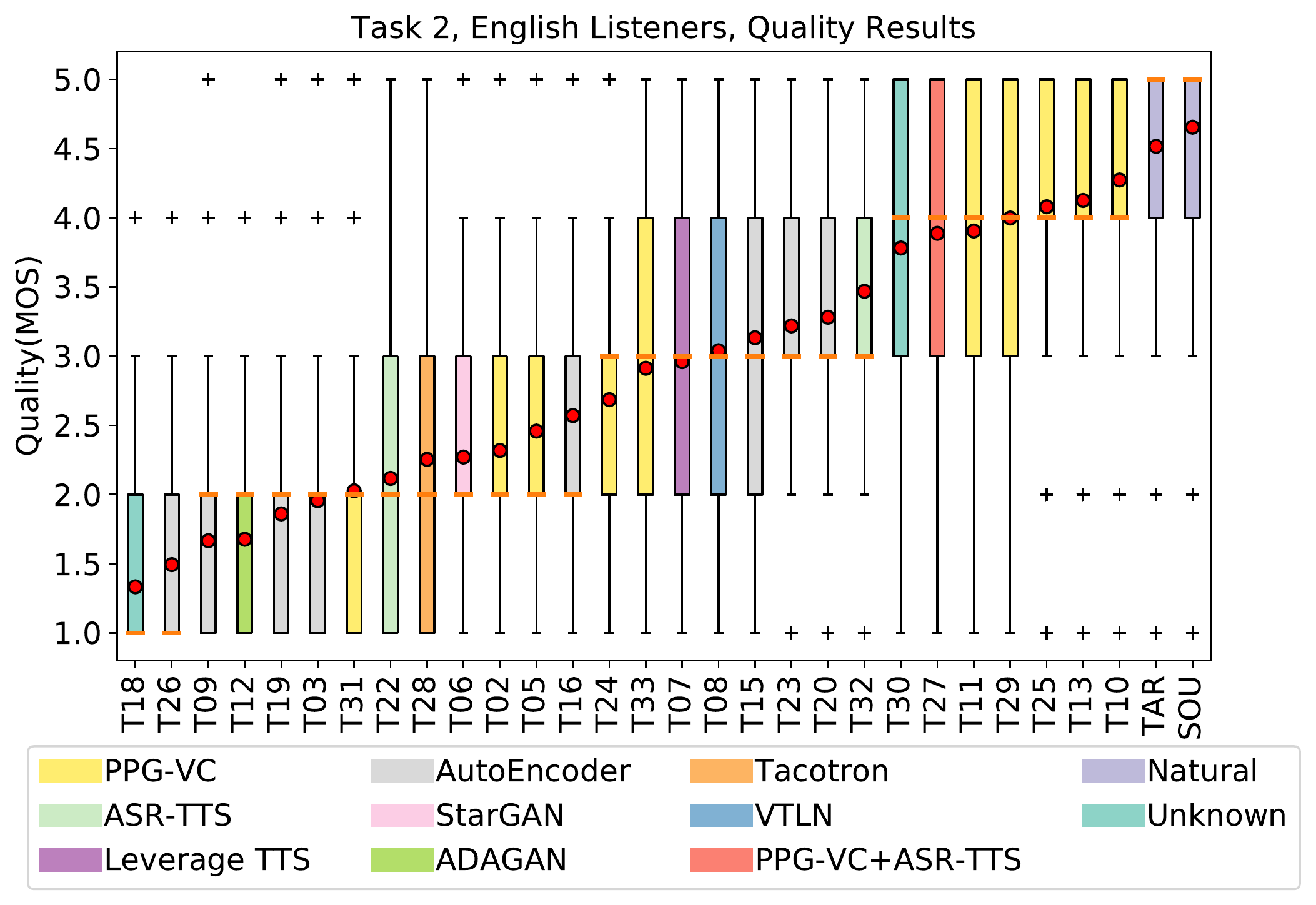} 
	\caption{Naturalness results for Task 2. MOS scores are arranged in accordance with their mean (red dot). Bars are colored on basis of feature conversion categories.	\label{fig:en_cross_score_qua}}
\end{figure*}

\subsubsection{Task 1 - Speaker similarity}

Figure~\ref{fig:en_intra_score_sim} shows the results for the speaker similarity evaluation for Task 1. The similarity percentage is defined as the added percentage of the same (not sure) and same (sure) scores for the system like in our previous challenges. The averaged similarity scores are also shown. Figure \ref{fig:en_intra_sim_sig} shows the significance groupings of the VC systems. Like naturalness, we used Wilcoxon signed-rank tests with Bonferroni correction ($\alpha$ = 0.01) for grouping systems that did not differ significantly from each other in Task 1.

\vspace{1mm}
\noindent 
\textbf{Comparison with T11:} From the figures, we can see that eight systems (T10, T22, T27, T13, T33, T23, T29, and T07) obtained the highest speaker similarity scores among the VC systems, and their improvements compared with T11 were statistically significant. We can therefore say that VC performance has improved within the past two years in terms of speaker similarity as well as naturalness. There were no significant differences among the eight best performing systems. 

\vspace{1mm}
\noindent 
\textbf{Comparison with humans:} According to Figure~\ref{fig:en_intra_sim_sig}, the similarity scores of the eight teams were not significantly different from the natural speech of the target speaker. In fact, over 90\% of the converted speech samples were judged to be the same as the target speakers by the listeners. \textit{In other words, the best performing systems achieved human-level speaker similarity, so the basic intra-lingual VC task has been solved in the sense of speaker similarity. We think that this is a historical achievement for VC research.} 

\vspace{1mm}
\noindent 
\textbf{PPG vs TTS:} We can see that, in addition to PPG-based approaches (T10 and T13), a combination of ASR and TTS systems (T22, T27, and T33) and hybrid TTS/VC systems (T07 and T23) also had equally good speaker similarity. 

Figure~\ref{fig:en_intra_scatter} shows a scatter plot matching naturalness and percentage of similarity to the target speaker for Task 1. We can see that many of the systems have trade-offs between naturalness and speaker similarity and that most have to improve either similarity or naturalness. T10 is again the closest to the natural speech of the target speaker.

\subsubsection{Task 2 - Naturalness}

\begin{figure*}[t]
	\centering
	\includegraphics[width=0.8\linewidth]{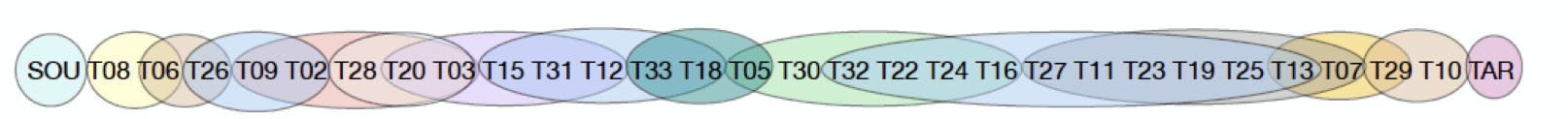} 
	\caption{ Groupings of systems that did not differ significantly from each other in terms of similarity for Task 2.}	\label{fig:en_cross_sim_sig}
\end{figure*}

Figure~\ref{fig:en_cross_score_qua} shows a boxplot and mean scores for the results of the naturalness evaluation for Task 2 (cross-lingual conversion task) when considering all L2 languages. Figure~\ref{fig:en_cross_sim_sig} shows the significance testing results. 

\vspace{1mm}
\noindent 
\textbf{Comparison with T11:} In a similar way to the basic intra-lingual VC results, we can first see that T10 obtained the highest MOS values among the VC systems and that it is significantly better than any of the other VC systems. We can also see that T13, T25, and T29 are also as good as T11 and not significantly different from T11. 

\vspace{1mm}
\noindent 
\textbf{Comparison with Task 1:} T10, however, suffered a drop in naturalness of 0.75 points in MOS score compared with the intra-lingual task. In fact, most teams who joined both tasks obtained a lower quality evaluation in Task 2, except for T08, T20, and T28. This clearly indicates an increase in the complexity of the cross-lingual conversion task.

\vspace{1mm}
\noindent 
\textbf{Comparison with human:} Moreover, it is interesting to observe that T10 achieved the highest scores and was not significantly different from the natural speech of the target speaker. 

\vspace{1mm}
\noindent 
\textbf{PPG vs TTS:} We can see that most of the best performing systems used PPG (T10, T13, T25). We can also see that VC systems using a combination of ASR and TTS systems (T27, T33, and T22) obtained lower scores. It seems that the PPG approach is more suitable than the combination of ASR and TTS in the case of the cross-lingual conversion task. 

\begin{figure*}[t]
	\centering
	\includegraphics[trim=1cm 0 1cm 9,clip,width=0.7\linewidth]{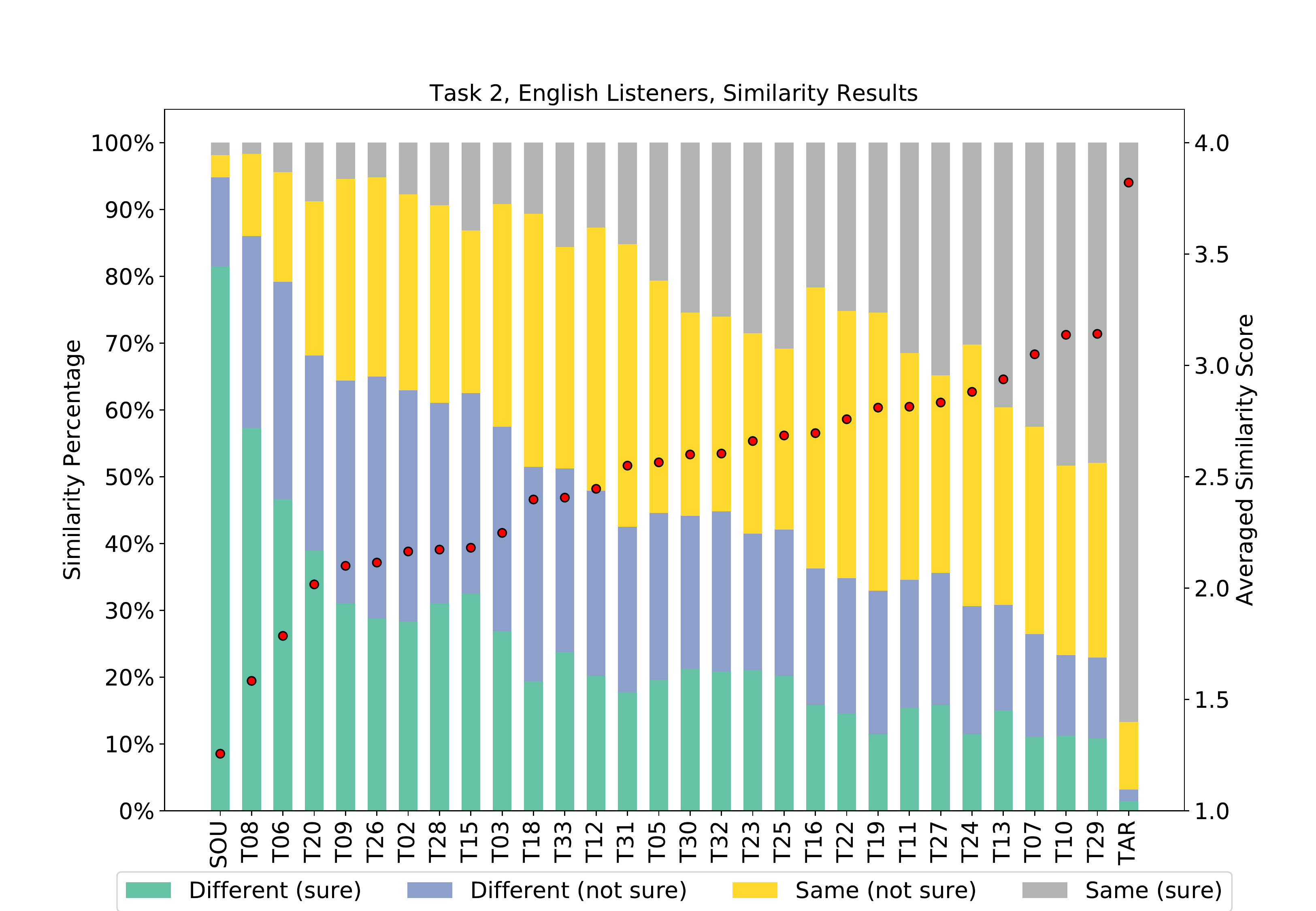} 
	\caption{Similarity results of target speaker for Task 2. Similarity scores are arranged in accordance with their mean value (red dot).	\label{fig:en_cross_score_sim}}
\end{figure*}

\begin{figure}[t]
	\centering
	\includegraphics[trim=1cm 1cm 1cm 1cm,clip,width=0.9\linewidth]{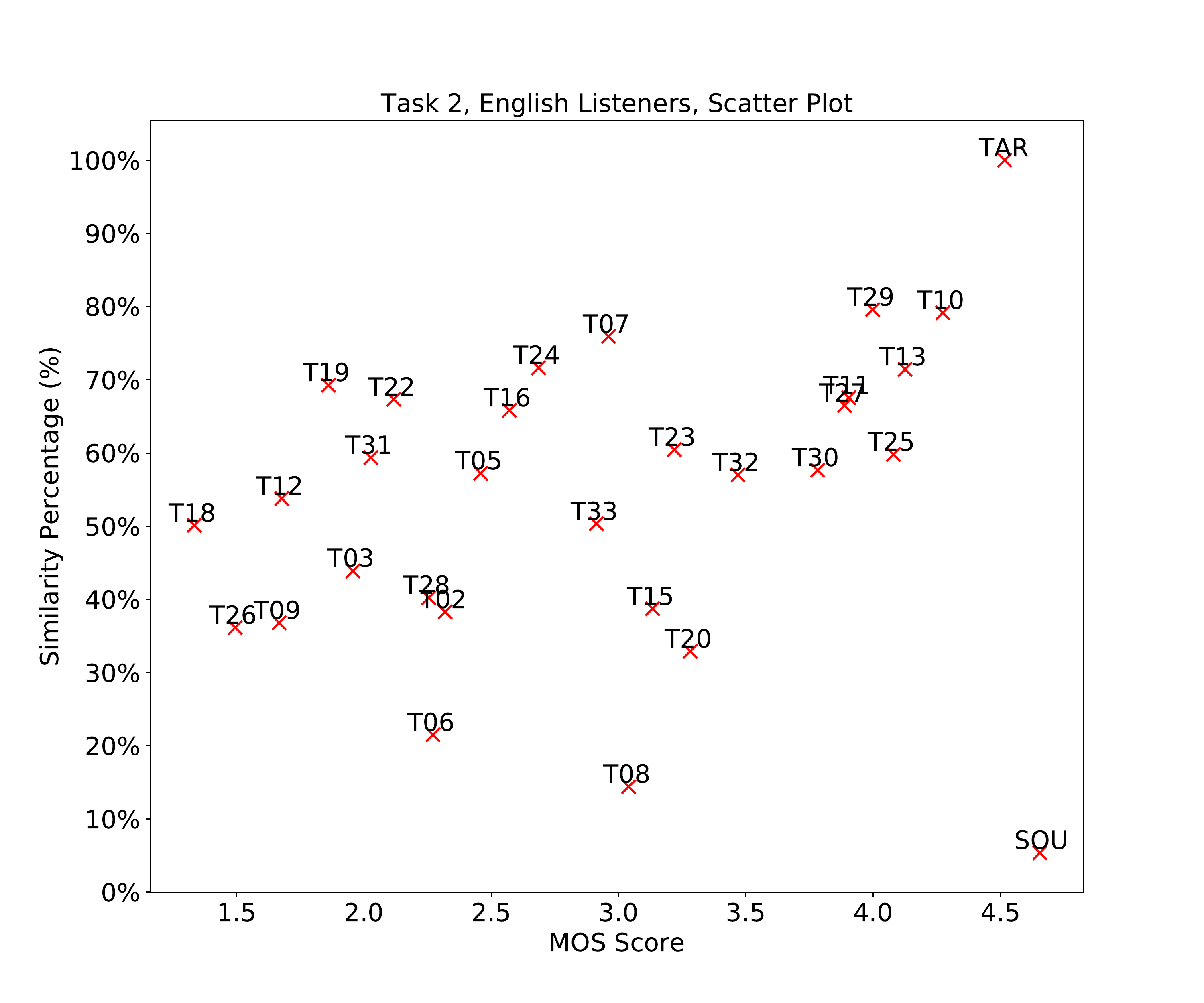} 
	\caption{Scatter plot matching naturalness and similarity scores to target speaker for cross-lingual conversion task when averaging all speaker pairs (y-axis is similarity percentage). \label{fig:en_cross_scatter}}
\end{figure}

\subsubsection{Task 2 - Speaker similarity}

Figure \ref{fig:en_cross_score_sim} shows the results of the speaker similarity evaluation for Task 2. Figure \ref{fig:en_cross_sim_sig} shows the significance groupings of the VC systems. 

\vspace{1mm}
\noindent 
\textbf{Comparison with T11:} From the figures, we can see that three systems (T10, T29, and T07) obtained encouraging results, and their speaker similarity scores were significantly better than T11. We can also see that T10 and T29 were not significantly different from each other, but only T10 was better than T07. 

\vspace{1mm}
\noindent 
\textbf{Comparison with humans and Task 1:} According to Figure~\ref{fig:en_cross_sim_sig}, all of the VC systems had much lower similarity scores than for natural speech, and the differences are statistically significant. Only less than 80\% of the converted speech samples were judged to be the same as target speakers by listeners. Compared with the basic intra-lingual VC task where the eight systems achieved human-level speaker similarity, the cross-lingual VC has room for improvement. 

\vspace{1mm}
\noindent 
\textbf{PPG vs TTS:} Again, it seems that the PPG approach (T10 and T29) and linguistic latent vector approach (T07) are more suitable than the combination of ASR and TTS in the case of the cross-lingual conversion task. 

Figure~\ref{fig:en_cross_scatter} shows a scatter plot matching naturalness and percentage of similarity to the target speaker for Task 2. We can see that, again, T10 and T29 were the closest to the actual natural speech of the target speaker and that there was an obvious gap between natural speech and VC systems.

\subsubsection{Summary of the listening tests}

From the above results of the listening tests for VCC 2020, we observed that VC methods have progressed rapidly. In particular, the speaker similarity scores of the best performing systems turned out to be as good as natural speech in the intra-lingual semi-parallel VC task, and not only PPG-based approaches but also a combination of ASR and TTS systems and hybrid TTS/VC systems also demonstrated good speaker conversion performance. However, we confirmed that none of them have achieved human-level naturalness yet for the intra-lingual semi-parallel VC task. One of the reasons we obtained different conclusions for naturalness and speaker similarity could be due to the abilities of human perception. It is known that humans do not have good speaker discrimination abilities and perform much worse than automatic speaker verification systems.

The cross-lingual conversion task is, as expected, a more difficult task, and the overall naturalness and similarity scores were lower than the intra-lingual conversion task. However, we observed encouraging results, and the MOS scores of the best systems were higher than 4.0. We also observed that the cascaded approach of ASR and TTS was not necessarily the best for the cross-lingual conversion task.

\section{Further analysis of VCC 2020 results}

Since this is the first large-scale listening test for cross-lingual VC, we carried out a few additional analyses to gain more insight into cross-lingual VC and the subjects' behavior in this section. 

One of our first questions was whether non-native listeners judge naturalness and speaker similarity in the same way as native subjects or not. How do they judge cross-lingual VC cases in particular? As we described earlier, we recruited both native speakers of English and non-native speakers of English (Japanese), and they completed identical listening tests in order to answer this question. 

\begin{figure}[t]
	\centering
	\includegraphics[trim=1cm 1cm 1cm 1cm,clip,width=1.0\linewidth]{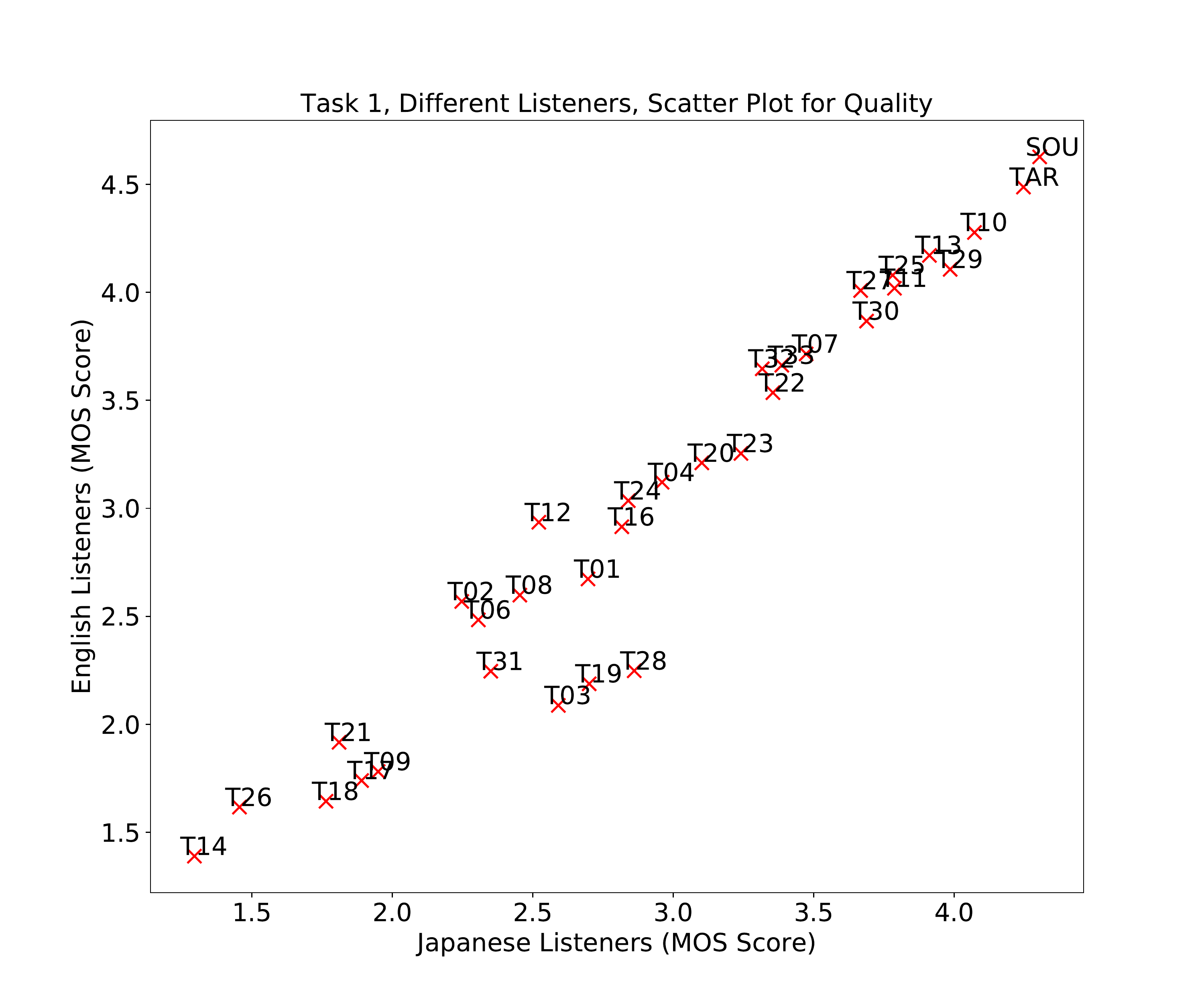}\\
	\includegraphics[trim=1cm 1cm 1cm 1cm,clip,width=1.0\linewidth]{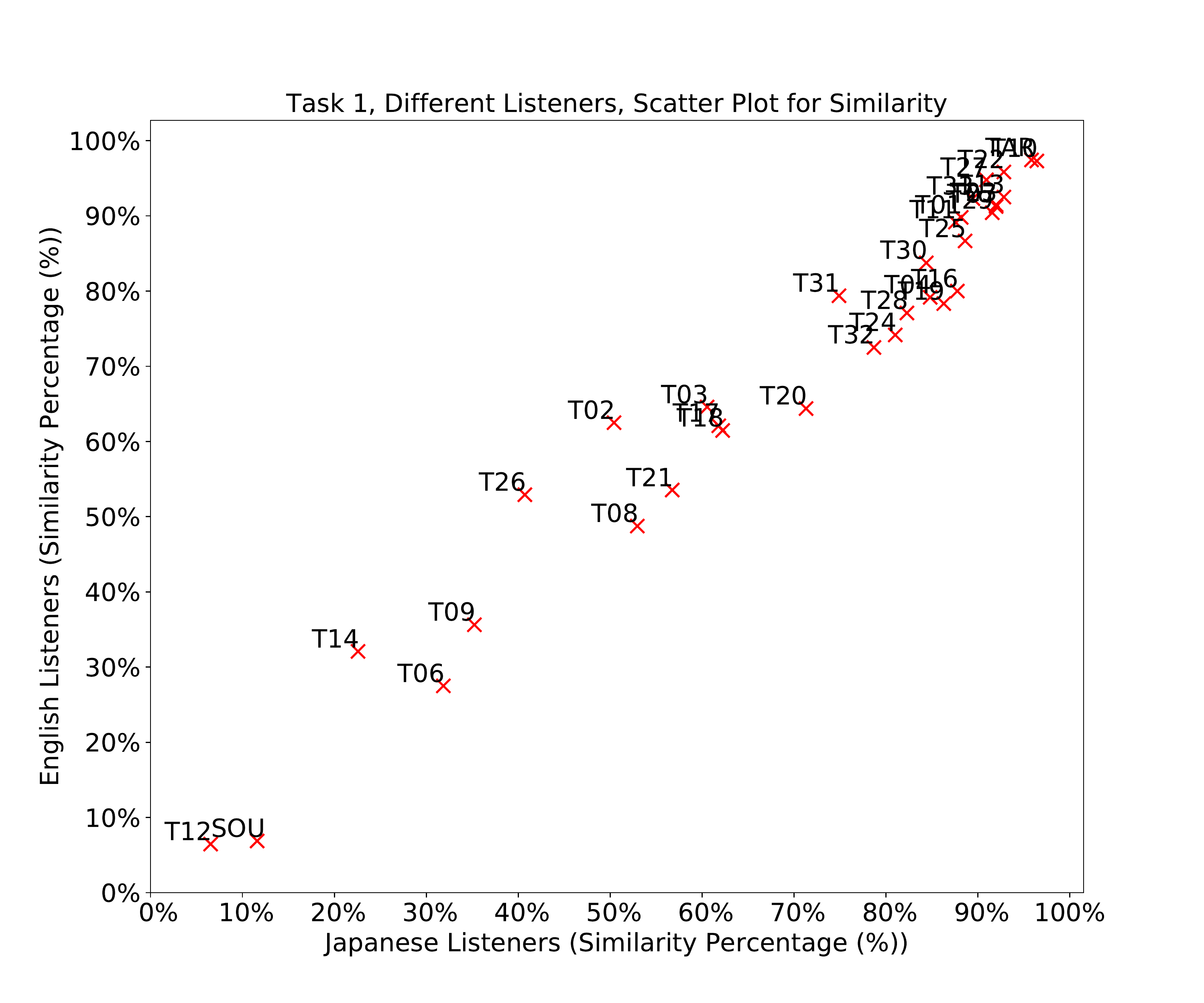} 
	\caption{Scatter plot matching naturalness scores (top) and similarity scores (bottom) of Japanese listeners and English listeners for Task 1. }\label{fig:jpen_qua_intra_scatter}
\end{figure}

\begin{figure}[t]
	\centering
	\includegraphics[trim=1cm 1cm 1cm 1cm,clip,width=1.0\linewidth]{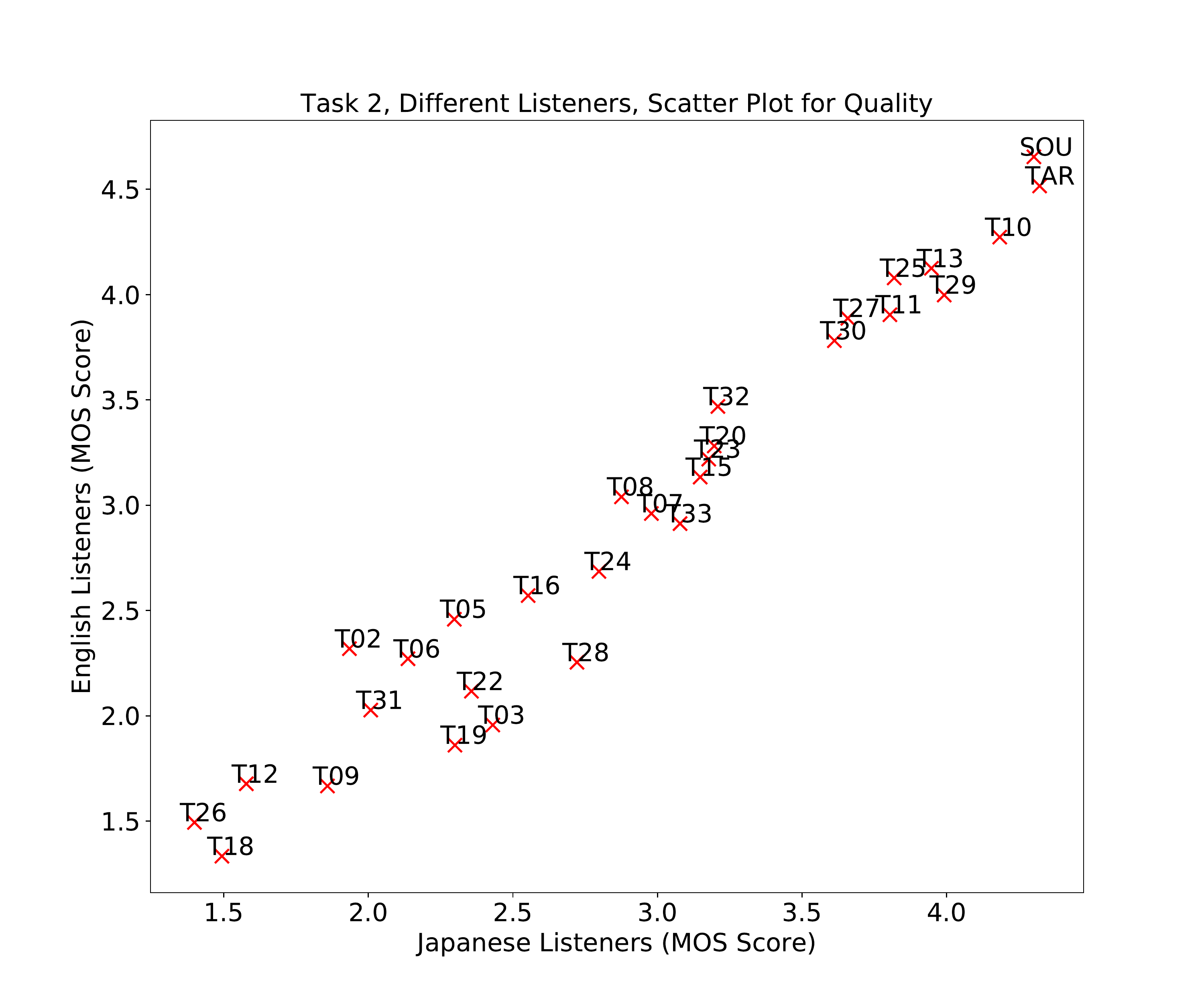} \\
	\includegraphics[trim=1cm 1cm 1cm 1cm,clip,width=1.0\linewidth]{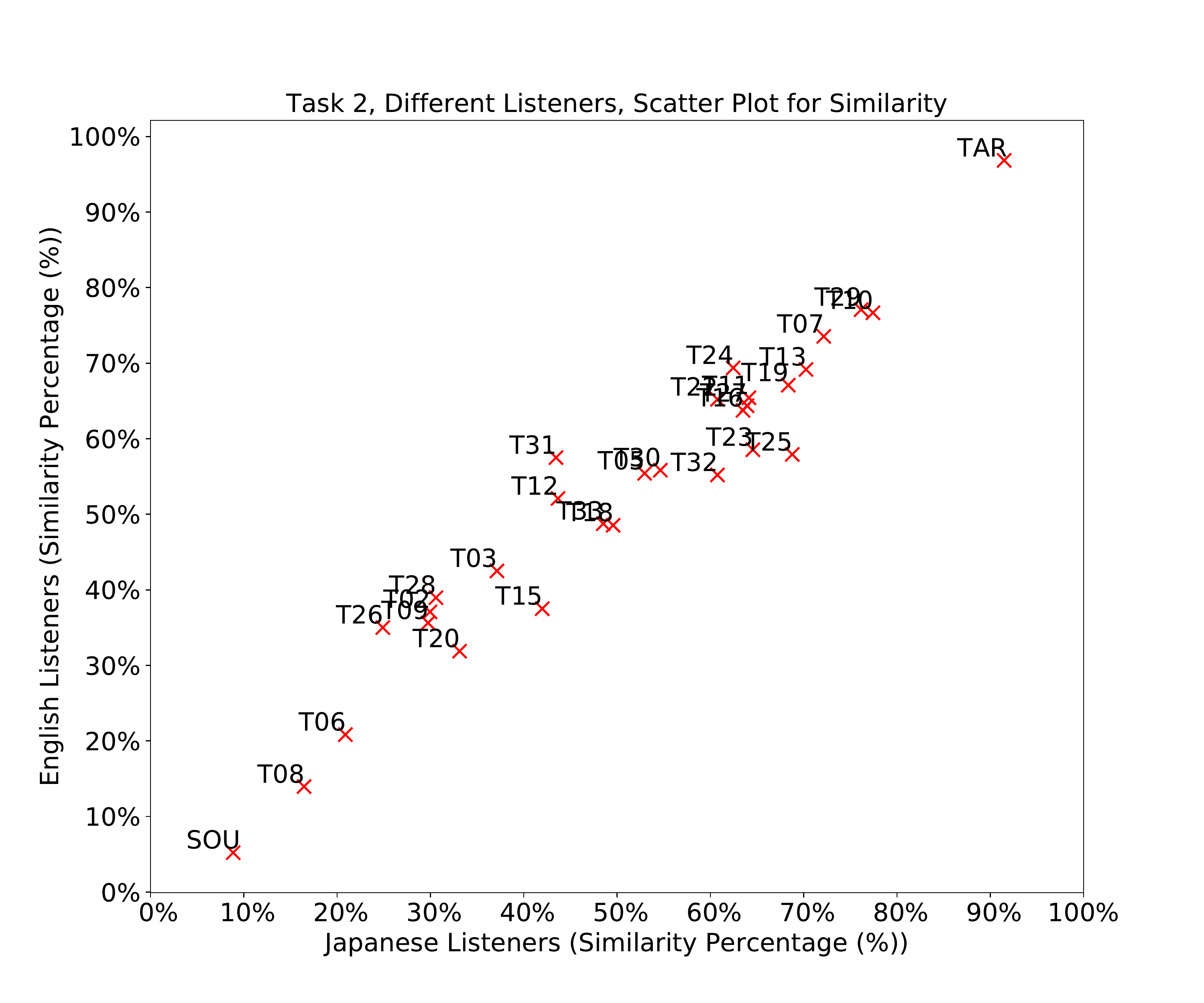} 
	\caption{Scatter plot matching naturalness scores (top) and similarity scores (bottom) of Japanese listeners and English listeners for Task 2.}\label{fig:JPEN_qua_cross_scatter}
\end{figure}

Second, we are also interested in finding out how subjects judge speaker similarity when they listen to reference audio in an L2 language. In the intra-lingual VC task, the reference audio is always in the same language as the input speech to be converted. However, in cross-lingual VC, the reference audio may be in a different language from that of input speech. It would be scientifically interesting to see whether subjects judge speaker similarity differently when they listen to reference audio spoken in a different language but uttered by the same target speaker. 

Third, the cross-lingual VC performance may differ according to the language of the target speaker. We therefore show a breakdown of cross-lingual VC performance per language of target speakers and analyze how the target speaker's language affected the performance.

\subsection{Do Japanese listeners judge naturalness and speaker similarity in the same way as English subjects?}

As we described earlier, the Japanese listeners also took the same listening tests as English listeners. We show detailed evaluation results (naturalness, speaker similarity, and their significant differences in Tasks 1 and 2) obtained with Japanese listeners in Appendix \ref{apped:Jpn}. Here, we discuss correlation and the difference in the evaluation results between these two kinds of listeners.

Figures \ref{fig:jpen_qua_intra_scatter} and \ref{fig:JPEN_qua_cross_scatter} show scatter plots matching the naturalness scores or speaker similarity scores of the Japanese listeners and English listeners for Task 1 or Task 2, respectively. As we can see from the figures, the Japanese listeners' judgements were well correlated with those of the English listeners in terms of both naturalness and speaker similarity. This is true for both Task 1 and Task 2. Their correlation values were 0.965, 0.984, 0.973, and 0.968, respectively. 

However, we can also see some minor differences between their judgements. For instance, English listeners gave lower scores than Japanese listeners for T28, T19, and T03 in both Tasks 1 and 2. After listening to the converted audio samples, we found that the three systems had very bad speech intelligibility. 

According to our results, although there were a few inconsistencies among the two listener groups, it seems acceptable to use non-native listeners to assess the performance of cross-lingual VC systems \textit{to some extent}.

\subsection{Do subjects judge speaker similarity differently when they listen to reference audio in L2 language?}

Next, we show the results to demonstrate how the subjects judged speaker similarity when they listened to the reference audio in either English or the L2 language. As described in Section 5.2, we presented reference audio recordings in English for three fifths of cases and presented audio recordings in L2 language uttered by the same target speakers for two fifths of cases by design, and, hence, we could compute speaker similarity scores separately according to the language of the reference audio. 
\begin{figure}[t]
	\centering
	\includegraphics[width=1.0\linewidth]{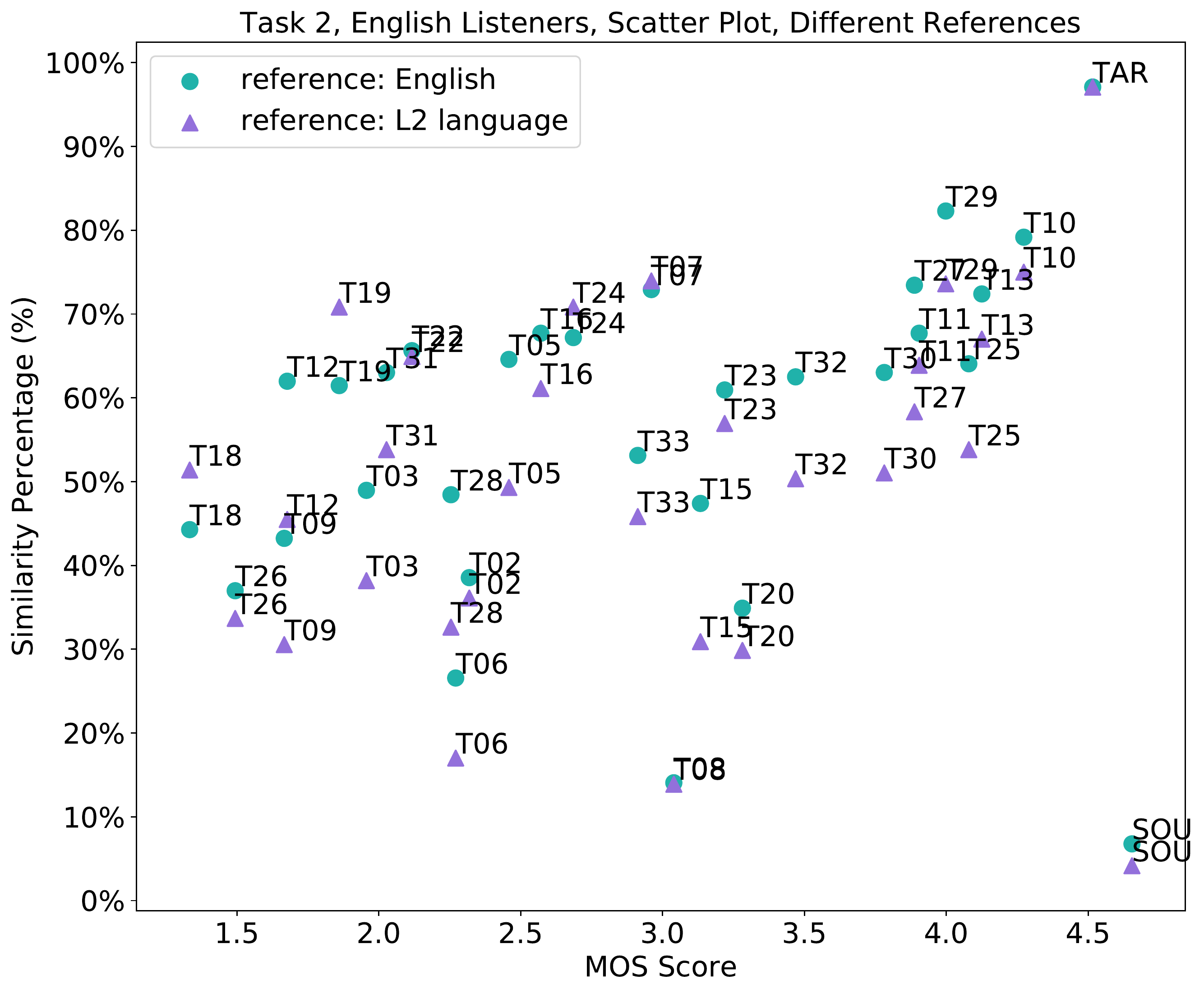} \\
	\includegraphics[width=1.0\linewidth]{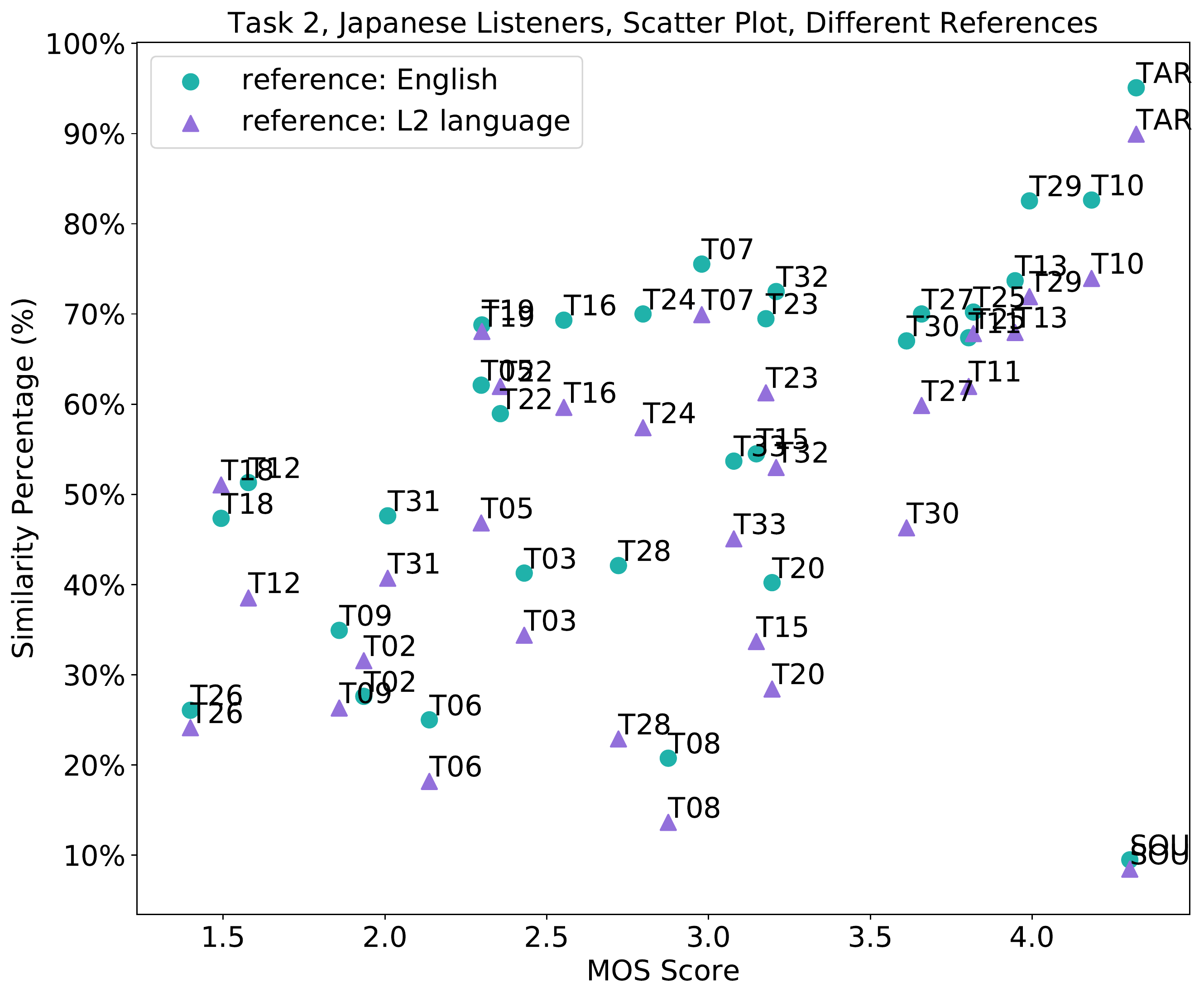} 
	\caption{Scatter plot matching naturalness and scores of similarity to target speaker for Task 2 when using different language audio as reference (y-axis is similarity percentage). Top part shows results for English listeners. Bottom part shows results for Japanese listeners.	\label{fig:en_l1l2_sim}}
\end{figure}

The top part of Figure~\ref{fig:en_l1l2_sim} shows a scatter plot including a breakdown of speaker-similarity evaluation results for English listeners, and the bottom part shows those for Japanese listeners. In the scatter plots, we used the same naturalness scores for both the English reference case and L2 language reference case for convenience. 

\begin{figure}[t]
	\centering
	\includegraphics[width=1.0\linewidth]{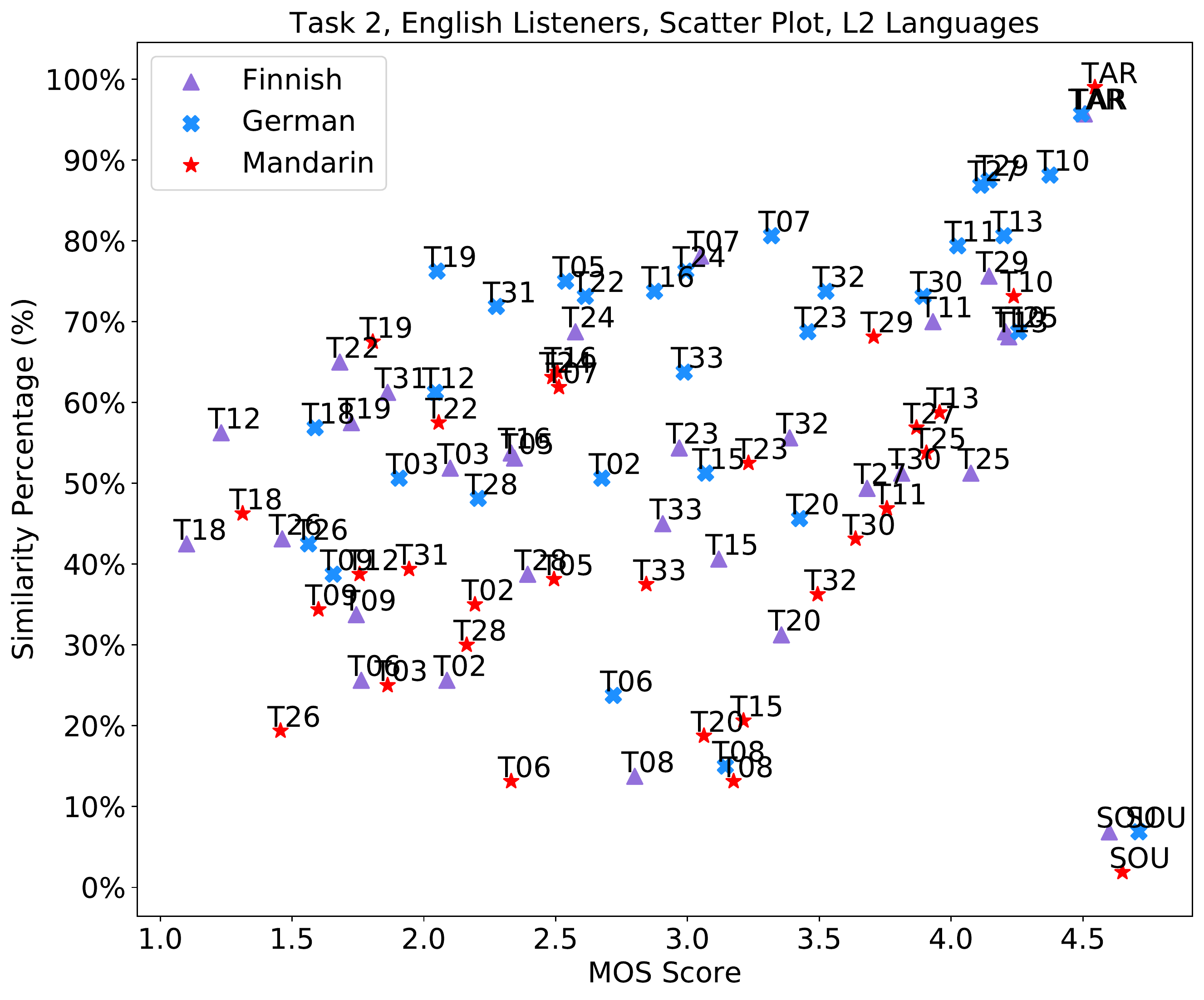} \\
	\includegraphics[width=1.0\linewidth]{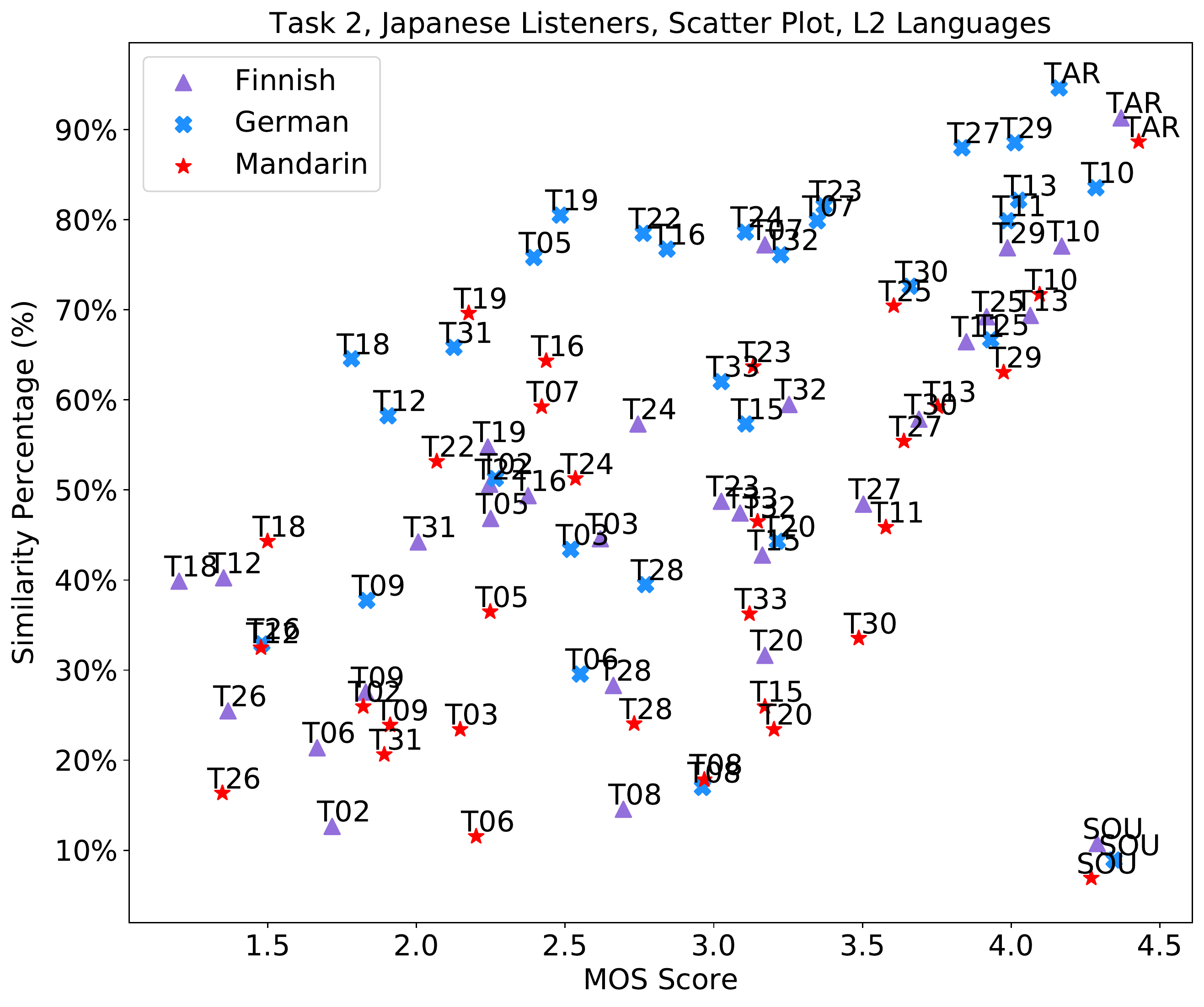} 
	\caption{Scatter plot matching naturalness and similarity scores of each L2 language to target speaker for Task 2 (y-axis is similarity percentage). Top part shows results for English listeners. Bottom part shows results for Japanese listeners.		\label{fig:en_l2langs_scatter}}
\end{figure}

We can see that subjects generally gave lower speaker similarity scores in the case of the L2 language reference even if the reference audio was uttered by the same target speaker. This would be because speaker verification across languages is not an easy task for humans as reported in \cite{WESTER2012781}, and, hence, subjects felt hesitation in choosing ``Same (sure).'' However, we can also see a few exceptions. T24, T19, and T18 had higher speaker similarity scores when the reference audio was in the L2 language. 

From this analysis, we could conclude that the choice of the language of the reference audio is very important. If the reference audio is different from that of input speech to be converted, speaker similarity scores given by subjects may become lower overall due to the language barrier. 

\begin{figure}[t]
	\centering
	\includegraphics[width=1.0\linewidth]{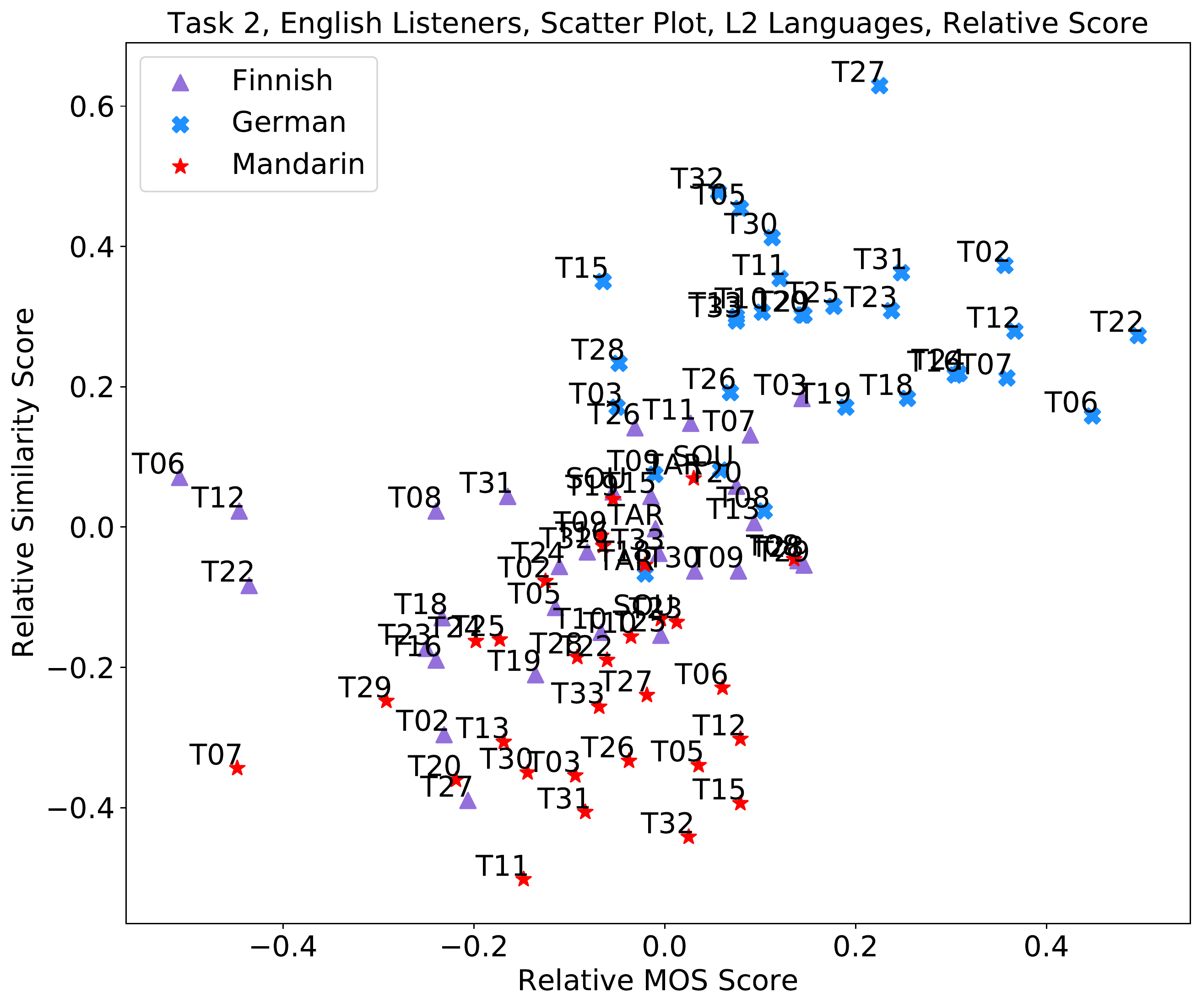} \\
	\includegraphics[width=1.0\linewidth]{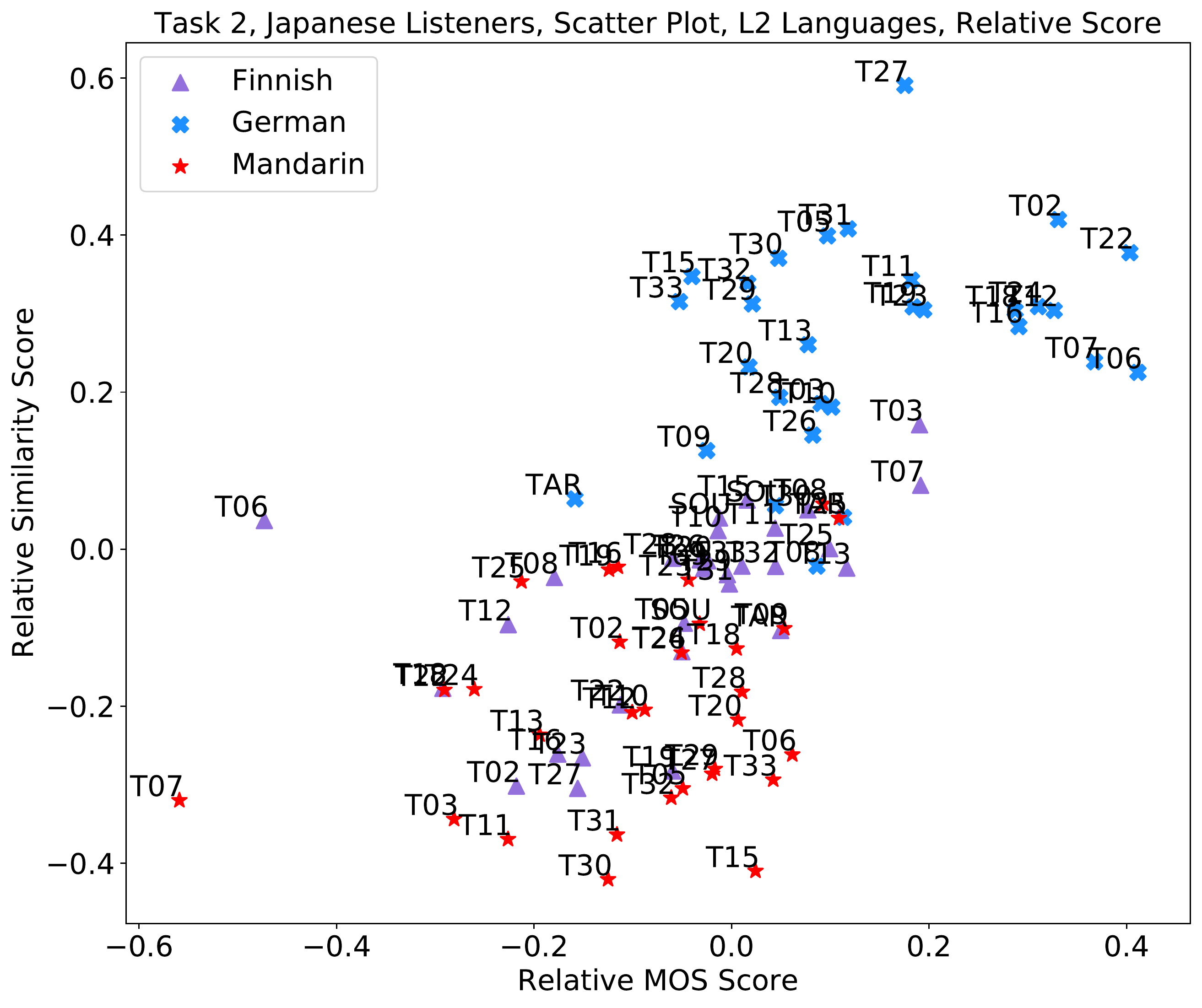} 
	\caption{Scatter plot matching relative naturalness and similarity scores of each L2 language to target speaker for Task 2 . Top part shows results for English listeners. Bottom part shows results for Japanese listeners.	\label{fig:en_l2langs_scatter_relative}}
\end{figure}

\subsection{Is cross-lingual VC performance affected by the language of target speakers?}

We can easily hypothesize that the cross-lingual VC performance may differ according to the language of the target speaker, but it is not clear whether it affects naturalness or speaker similarity or both. Here, we therefore show a breakdown of the cross-lingual VC performance per target-speaker language in Figure \ref{fig:en_l2langs_scatter}. The top and bottom parts show the results for English and Japanese listeners, respectively. To show relative differences among the languages, we also computed the relative scores of each language to the mean and plotted the results in Figure~\ref{fig:en_l2langs_scatter_relative}.

From the figure, we can first see that the language of the target speakers affected both the speaker similarity and naturalness of the VC systems. We can also observe that most of the VC systems had the highest MOS and similarity scores for German target speakers and lowest similarity scores for Mandarin speakers. This may be partially due to linguistic distances to English, but this requires further investigation. 

\section{Conclusion}

The voice conversion challenge is a bi-annual scientific event held to compare and understand different VC systems built on a common dataset. In 2020, we organized the third edition of the challenge and constructed and distributed a new database for two tasks, intra-lingual semi-parallel and cross-lingual voice conversion. The participants were given two months and two weeks to build VC systems, and we received a total of 33 submissions, including 3 baselines built on the database. From the results of crowd-sourced listening tests, we saw that VC methods have progressed rapidly thanks to advanced deep learning methods. In particular, the speaker similarity scores of several systems turned out to be as high as target speakers in the intra-lingual semi-parallel VC task. However, we confirmed that none of them have achieved human-level naturalness yet for the same task. The cross-lingual conversion task is, as expected, a more difficult task, and the overall naturalness and similarity scores were lower than the intra-lingual conversion task. However, we observed encouraging results, and the MOS scores of the best systems were higher than 4.0.

We also provided a few additional analysis results to aid in understanding cross-lingual voice conversion better. We tried to answer three questions and showed our insights: 1) Do Japanese listeners judge naturalness and speaker similarity in the same way as English subjects? 2) Do subjects judge speaker similarity differently when they listen to reference audio in L2 language? and 3) Is cross-lingual VC performance affected by the language of target speakers? The insights could help us to improve and evaluate cross-lingual voice conversion in the future. 
\section{Acknowledgements}
This work was partially supported by JST CREST Grants (JPMJCR18A6, VoicePersonae project, and JPMJCR19A3, CoAugmentation project), Japan, MEXT KAKENHI Grants (16H06302, 17H04687, 17H06101, 18H04120, 18H04112, 18KT0051, 19K24373), Japan, the National Natural Science Foundation of China (Grant No. 61871358) and Programmatic Grant No. A1687b0033 from the Singapore Government’s Research, Innovation and Enterprise 2020 plan (Advanced Manufacturing and Engineering domain), and Academy of Finland (project no. 309629)

\balance
\bibliographystyle{IEEEtran}
\bibliography{main}

\begin{thebibliography}{10}
\providecommand{\url}[1]{#1}
\csname url@samestyle\endcsname
\providecommand{\newblock}{\relax}
\providecommand{\bibinfo}[2]{#2}
\providecommand{\BIBentrySTDinterwordspacing}{\spaceskip=0pt\relax}
\providecommand{\BIBentryALTinterwordstretchfactor}{4}
\providecommand{\BIBentryALTinterwordspacing}{\spaceskip=\fontdimen2\font plus
\BIBentryALTinterwordstretchfactor\fontdimen3\font minus
  \fontdimen4\font\relax}
\providecommand{\BIBforeignlanguage}[2]{{%
\expandafter\ifx\csname l@#1\endcsname\relax
\typeout{** WARNING: IEEEtran.bst: No hyphenation pattern has been}%
\typeout{** loaded for the language `#1'. Using the pattern for}%
\typeout{** the default language instead.}%
\else
\language=\csname l@#1\endcsname
\fi
#2}}
\providecommand{\BIBdecl}{\relax}
\BIBdecl

\bibitem{Kain2007dysarthric}
A.~Kain, J.-P. Hosom, X.~Niu, J.~Santen, M.~Fried-Oken, and J.~Staehely,
  ``Improving the intelligibility of dysarthric speech,'' \emph{Speech
  Communication}, vol.~49, no.~09, pp. 743--759, 2007.

\bibitem{Felps2009accent}
D.~Felps, H.~Bortfeld, and R.~Gutierrez-Osuna, ``Foreign accent conversion in
  computer assisted pronunciation training,'' \emph{Speech communication},
  vol.~51, no.~10, pp. 920--932, 2009.

\bibitem{Turk2010expressive}
O.~T\"{u}rk and M.~Schr\"{o}der, ``Evaluation of expressive speech synthesis
  with voice conversion and copy resynthesis techniques,'' \emph{IEEE/ACM
  Transactions on Audio, Speech, and Language Processing}, vol.~18, no.~5, pp.
  965--973, 2010.

\bibitem{Villavicencio2010singing}
F.~Villavicencio and J.~Bonada, ``Applying voice conversion to concatenative
  singing-voice synthesis.'' in \emph{Proc. Interspeech}, 2010, pp. 2162--2165.

\bibitem{Toda2012silent}
T.~Toda, M.~Nakagiri, and K.~Shikano, ``Statistical voice conversion techniques
  for body-conducted unvoiced speech enhancement,'' \emph{IEEE Transactions on
  Audio, Speech, and Language Processing}, vol.~20, no.~9, pp. 2505--2517,
  2012.

\bibitem{toda2016voice}
T.~Toda, L.-H. Chen, D.~Saito, F.~Villavicencio, M.~Wester, Z.~Wu, and
  J.~Yamagishi, ``The voice conversion challenge 2016.'' in \emph{Interspeech},
  2016, pp. 1632--1636.

\bibitem{Lorenzo-Trueba2018}
\BIBentryALTinterwordspacing
J.~Lorenzo-Trueba, J.~Yamagishi, T.~Toda, D.~Saito, F.~Villavicencio,
  T.~Kinnunen, and Z.~Ling, ``The voice conversion challenge 2018: Promoting
  development of parallel and nonparallel methods,'' in \emph{Proc. Odyssey
  2018 The Speaker and Language Recognition Workshop}, 2018, pp. 195--202.
  [Online]. Available: \url{http://dx.doi.org/10.21437/Odyssey.2018-28}
\BIBentrySTDinterwordspacing

\bibitem{wester2016analysis}
M.~Wester, Z.~Wu, and J.~Yamagishi, ``Analysis of the voice conversion
  challenge 2016 evaluation results.'' in \emph{Interspeech}, 2016, pp.
  1637--1641.

\bibitem{emime}
M.~Wester, ``The {EMIME} bilingual database,'' The University of Edinburgh,
  Tech. Rep., 2010.

\bibitem{sisman2020overview}
B.~Sisman, J.~Yamagishi, S.~King, and H.~Li, ``An overview of voice conversion
  and its challenges: From statistical modeling to deep learning,'' \emph{arXiv
  preprint arXiv:2008.03648}, 2020.

\bibitem{tobing2019non}
P.~L. Tobing, Y.-C. Wu, T.~Hayashi, K.~Kobayashi, and T.~Toda, ``Non-parallel
  voice conversion with cyclic variational autoencoder,'' in \emph{Proc.
  Interspeech}, 2019, pp. 674--678.

\bibitem{9053512}
T.~{Hayashi}, R.~{Yamamoto}, K.~{Inoue}, T.~{Yoshimura}, S.~{Watanabe},
  T.~{Toda}, K.~{Takeda}, Y.~{Zhang}, and X.~{Tan}, ``Espnet-tts: Unified,
  reproducible, and integratable open source end-to-end text-to-speech
  toolkit,'' in \emph{ICASSP 2020 - 2020 IEEE International Conference on
  Acoustics, Speech and Signal Processing (ICASSP)}, 2020, pp. 7654--7658.

\bibitem{sun2016phonetic}
L.~Sun, K.~Li, H.~Wang, S.~Kang, and H.~Meng, ``Phonetic posteriorgrams for
  many-to-one voice conversion without parallel data training,'' in
  \emph{International Conference on Multimedia and Expo (ICME)}.\hskip 1em plus
  0.5em minus 0.4em\relax IEEE, 2016, pp. 1--6.

\bibitem{tian2018average}
X.~Tian, J.~Wang, H.~Xu, E.~S. Chng, and H.~Li, ``Average modeling approach to
  voice conversion with non-parallel data.'' in \emph{Odyssey}, vol. 2018,
  2018, pp. 227--232.

\bibitem{Liu2018}
\BIBentryALTinterwordspacing
L.-J. Liu, Z.-H. Ling, Y.~Jiang, M.~Zhou, and L.-R. Dai, ``{WaveNet} vocoder
  with limited training data for voice conversion,'' in \emph{Proc.
  Interspeech}, 2018, pp. 1983--1987. [Online]. Available:
  \url{http://dx.doi.org/10.21437/Interspeech.2018-1190}
\BIBentrySTDinterwordspacing

\bibitem{qian2019autovc}
K.~Qian, Y.~Zhang, S.~Chang, X.~Yang, and M.~Hasegawa-Johnson, ``Autovc:
  Zero-shot voice style transfer with only autoencoder loss,'' \emph{arXiv
  preprint arXiv:1905.05879}, 2019.

\bibitem{van2017neural}
A.~Van Den~Oord, O.~Vinyals \emph{et~al.}, ``Neural discrete representation
  learning,'' in \emph{Advances in Neural Information Processing Systems},
  2017, pp. 6306--6315.

\bibitem{chou2019one}
J.-c. Chou, C.-c. Yeh, and H.-y. Lee, ``One-shot voice conversion by separating
  speaker and content representations with instance normalization,''
  \emph{arXiv preprint arXiv:1904.05742}, 2019.

\bibitem{hsu2017voice}
C.-C. Hsu, H.-T. Hwang, Y.-C. Wu, Y.~Tsao, and H.-M. Wang, ``Voice conversion
  from unaligned corpora using variational autoencoding wasserstein generative
  adversarial networks,'' \emph{arXiv preprint arXiv:1704.00849}, 2017.

\bibitem{dehak2010front}
N.~Dehak, P.~J. Kenny, R.~Dehak, P.~Dumouchel, and P.~Ouellet, ``Front-end
  factor analysis for speaker verification,'' \emph{IEEE Transactions on Audio,
  Speech, and Language Processing}, vol.~19, no.~4, pp. 788--798, 2010.

\bibitem{snyder2018x}
D.~Snyder, D.~Garcia-Romero, G.~Sell, D.~Povey, and S.~Khudanpur, ``X-vectors:
  Robust dnn embeddings for speaker recognition,'' in \emph{International
  Conference on Acoustics, Speech and Signal Processing (ICASSP)}.\hskip 1em
  plus 0.5em minus 0.4em\relax IEEE, 2018, pp. 5329--5333.

\bibitem{zhou2019cross}
Y.~Zhou, X.~Tian, H.~Xu, R.~K. Das, and H.~Li, ``Cross-lingual voice conversion
  with bilingual phonetic posteriorgram and average modeling,'' in
  \emph{International Conference on Acoustics, Speech and Signal Processing
  (ICASSP)}.\hskip 1em plus 0.5em minus 0.4em\relax IEEE, 2019, pp. 6790--6794.

\bibitem{wang2017tacotron}
Y.~Wang, R.~Skerry-Ryan, D.~Stanton, Y.~Wu, R.~J. Weiss, N.~Jaitly, Z.~Yang,
  Y.~Xiao, Z.~Chen, S.~Bengio \emph{et~al.}, ``Tacotron: Towards end-to-end
  speech synthesis,'' \emph{arXiv preprint arXiv:1703.10135}, 2017.

\bibitem{vaswani2017attention}
A.~Vaswani, N.~Shazeer, N.~Parmar, J.~Uszkoreit, L.~Jones, A.~N. Gomez,
  {\L}.~Kaiser, and I.~Polosukhin, ``Attention is all you need,'' in
  \emph{Advances in neural information processing systems}, 2017, pp.
  5998--6008.

\bibitem{li2019neural}
N.~Li, S.~Liu, Y.~Liu, S.~Zhao, and M.~Liu, ``Neural speech synthesis with
  transformer network,'' in \emph{Proceedings of the AAAI Conference on
  Artificial Intelligence}, vol.~33, 2019, pp. 6706--6713.

\bibitem{kingma2016improved}
D.~P. Kingma, T.~Salimans, R.~Jozefowicz, X.~Chen, I.~Sutskever, and
  M.~Welling, ``Improved variational inference with inverse autoregressive
  flow,'' in \emph{Advances in neural information processing systems}, 2016,
  pp. 4743--4751.

\bibitem{biadsy2019parrotron}
F.~Biadsy, R.~J. Weiss, P.~J. Moreno, D.~Kanevsky, and Y.~Jia, ``Parrotron: An
  end-to-end speech-to-speech conversion model and its applications to
  hearing-impaired speech and speech separation,'' \emph{arXiv preprint
  arXiv:1904.04169}, 2019.

\bibitem{park2020cotatron}
S.-w. Park, D.-y. Kim, and M.-c. Joe, ``Cotatron: Transcription-guided speech
  encoder for any-to-many voice conversion without parallel data,'' \emph{arXiv
  preprint arXiv:2005.03295}, 2020.

\bibitem{huang2019voice}
W.-C. Huang, T.~Hayashi, Y.-C. Wu, H.~Kameoka, and T.~Toda, ``Voice transformer
  network: Sequence-to-sequence voice conversion using transformer with
  text-to-speech pretraining,'' \emph{arXiv preprint arXiv:1912.06813}, 2019.

\bibitem{luong2020nautilus}
H.-T. Luong and J.~Yamagishi, ``{NAUTILUS}: a versatile voice cloning system,''
  \emph{arXiv preprint arXiv:2005.11004}, 2020.

\bibitem{zhu2017unpaired}
J.-Y. Zhu, T.~Park, P.~Isola, and A.~A. Efros, ``Unpaired image-to-image
  translation using cycle-consistent adversarial networks,'' in
  \emph{Proceedings of the IEEE international conference on computer vision},
  2017, pp. 2223--2232.

\bibitem{kaneko2018cyclegan}
T.~Kaneko and H.~Kameoka, ``Cyclegan-vc: Non-parallel voice conversion using
  cycle-consistent adversarial networks,'' in \emph{26th European Signal
  Processing Conference (EUSIPCO)}.\hskip 1em plus 0.5em minus 0.4em\relax
  IEEE, 2018, pp. 2100--2104.

\bibitem{kameoka2018stargan}
H.~Kameoka, T.~Kaneko, K.~Tanaka, and N.~Hojo, ``Stargan-vc: Non-parallel
  many-to-many voice conversion using star generative adversarial networks,''
  in \emph{Spoken Language Technology Workshop (SLT)}.\hskip 1em plus 0.5em
  minus 0.4em\relax IEEE, 2018, pp. 266--273.

\bibitem{patel2019adagan}
M.~Patel, M.~Purohit, M.~Parmar, N.~J. Shah, and H.~A. Patil, ``Adagan:
  Adaptive gan for many-to-many non-parallel voice conversion,'' 2019.

\bibitem{sundermann2003vtln}
D.~Sundermann and H.~Ney, ``{VTLN}-based voice conversion,'' in
  \emph{Proceedings of the 3rd IEEE International Symposium on Signal
  Processing and Information Technology}.\hskip 1em plus 0.5em minus
  0.4em\relax IEEE, 2003, pp. 556--559.

\bibitem{kobayashi2018statistical}
K.~Kobayashi, T.~Toda, and S.~Nakamura, ``Intra-gender statistical singing
  voice conversion with direct waveform modification using log-spectral
  differential,'' \emph{Speech Communication}, vol.~99, pp. 211--220, 2018.

\bibitem{erro2009inca}
D.~Erro, A.~Moreno, and A.~Bonafonte, ``{INCA} algorithm for training voice
  conversion systems from nonparallel corpora,'' \emph{IEEE Transactions on
  Audio, Speech, and Language Processing}, vol.~18, no.~5, pp. 944--953, 2009.

\bibitem{oord2016wavenet}
A.~v.~d. Oord, S.~Dieleman, H.~Zen, K.~Simonyan, O.~Vinyals, A.~Graves,
  N.~Kalchbrenner, A.~Senior, and K.~Kavukcuoglu, ``{WaveNet}: A generative
  model for raw audio,'' \emph{arXiv preprint arXiv:1609.03499}, 2016.

\bibitem{kalchbrenner2018efficient}
N.~Kalchbrenner, E.~Elsen, K.~Simonyan, S.~Noury, N.~Casagrande, E.~Lockhart,
  F.~Stimberg, A.~v.~d. Oord, S.~Dieleman, and K.~Kavukcuoglu, ``Efficient
  neural audio synthesis,'' \emph{arXiv preprint arXiv:1802.08435}, 2018.

\bibitem{valin2019lpcnet}
J.-M. Valin and J.~Skoglund, ``{LPCNet}: Improving neural speech synthesis
  through linear prediction,'' in \emph{International Conference on Acoustics,
  Speech and Signal Processing (ICASSP)}.\hskip 1em plus 0.5em minus
  0.4em\relax IEEE, 2019, pp. 5891--5895.

\bibitem{yamamoto2020parallel}
R.~Yamamoto, E.~Song, and J.-M. Kim, ``Parallel {WaveGAN}: A fast waveform
  generation model based on generative adversarial networks with
  multi-resolution spectrogram,'' in \emph{ICASSP 2020-2020 IEEE International
  Conference on Acoustics, Speech and Signal Processing (ICASSP)}.\hskip 1em
  plus 0.5em minus 0.4em\relax IEEE, 2020, pp. 6199--6203.

\bibitem{prenger2019waveglow}
R.~Prenger, R.~Valle, and B.~Catanzaro, ``Waveglow: A flow-based generative
  network for speech synthesis,'' in \emph{International Conference on
  Acoustics, Speech and Signal Processing (ICASSP)}.\hskip 1em plus 0.5em minus
  0.4em\relax IEEE, 2019, pp. 3617--3621.

\bibitem{kumar2019melgan}
K.~Kumar, R.~Kumar, T.~de~Boissiere, L.~Gestin, W.~Z. Teoh, J.~Sotelo,
  A.~de~Br{\'e}bisson, Y.~Bengio, and A.~C. Courville, ``{MelGAN}: Generative
  adversarial networks for conditional waveform synthesis,'' in \emph{Advances
  in Neural Information Processing Systems}, 2019, pp. 14\,910--14\,921.

\bibitem{wang2019neural}
X.~Wang, S.~Takaki, and J.~Yamagishi, ``Neural source-filter waveform models
  for statistical parametric speech synthesis,'' \emph{IEEE/ACM Transactions on
  Audio, Speech, and Language Processing}, vol.~28, pp. 402--415, 2019.

\bibitem{griffin1984signal}
D.~Griffin and J.~Lim, ``Signal estimation from modified short-time fourier
  transform,'' \emph{IEEE Transactions on Acoustics, Speech, and Signal
  Processing}, vol.~32, no.~2, pp. 236--243, 1984.

\bibitem{morise2016world}
M.~Morise, F.~Yokomori, and K.~Ozawa, ``{WORLD}: a vocoder-based high-quality
  speech synthesis system for real-time applications,'' \emph{IEICE
  TRANSACTIONS on Information and Systems}, vol.~99, no.~7, pp. 1877--1884,
  2016.

\bibitem{erro2013harmonics}
D.~Erro, I.~Sainz, E.~Navas, and I.~Hernaez, ``Harmonics plus noise model based
  vocoder for statistical parametric speech synthesis,'' \emph{IEEE Journal of
  Selected Topics in Signal Processing}, vol.~8, no.~2, pp. 184--194, 2013.

\bibitem{vcc2020objective}
T.~Kinnunen, R.~K. Das, Z.~Ling, Y.~Zhao, J.~Yamagishi, X.~Tian, W.-C. Huang,
  and T.~Toda, ``Objective assessment of voice conversion challenge 2020
  submissions,'' in \emph{ISCA Joint Workshop for the Blizzard Challenge and
  Voice Conversion Challenge 2020}.\hskip 1em plus 0.5em minus 0.4em\relax
  ISCA, 2020, pp. XXX--XXX.

\bibitem{WESTER2012781}
\BIBentryALTinterwordspacing
M.~Wester, ``Talker discrimination across languages,'' \emph{Speech
  Communication}, vol.~54, no.~6, pp. 781 -- 790, 2012. [Online]. Available:
  \url{http://www.sciencedirect.com/science/article/pii/S0167639312000131}
\BIBentrySTDinterwordspacing

\end{thebibliography}

\appendix

\onecolumn

\section{Vocoder type comparisons}
\label{apped:vocoder} 

\begin{figure}[h]
	\centering
\includegraphics[width=0.65\linewidth]{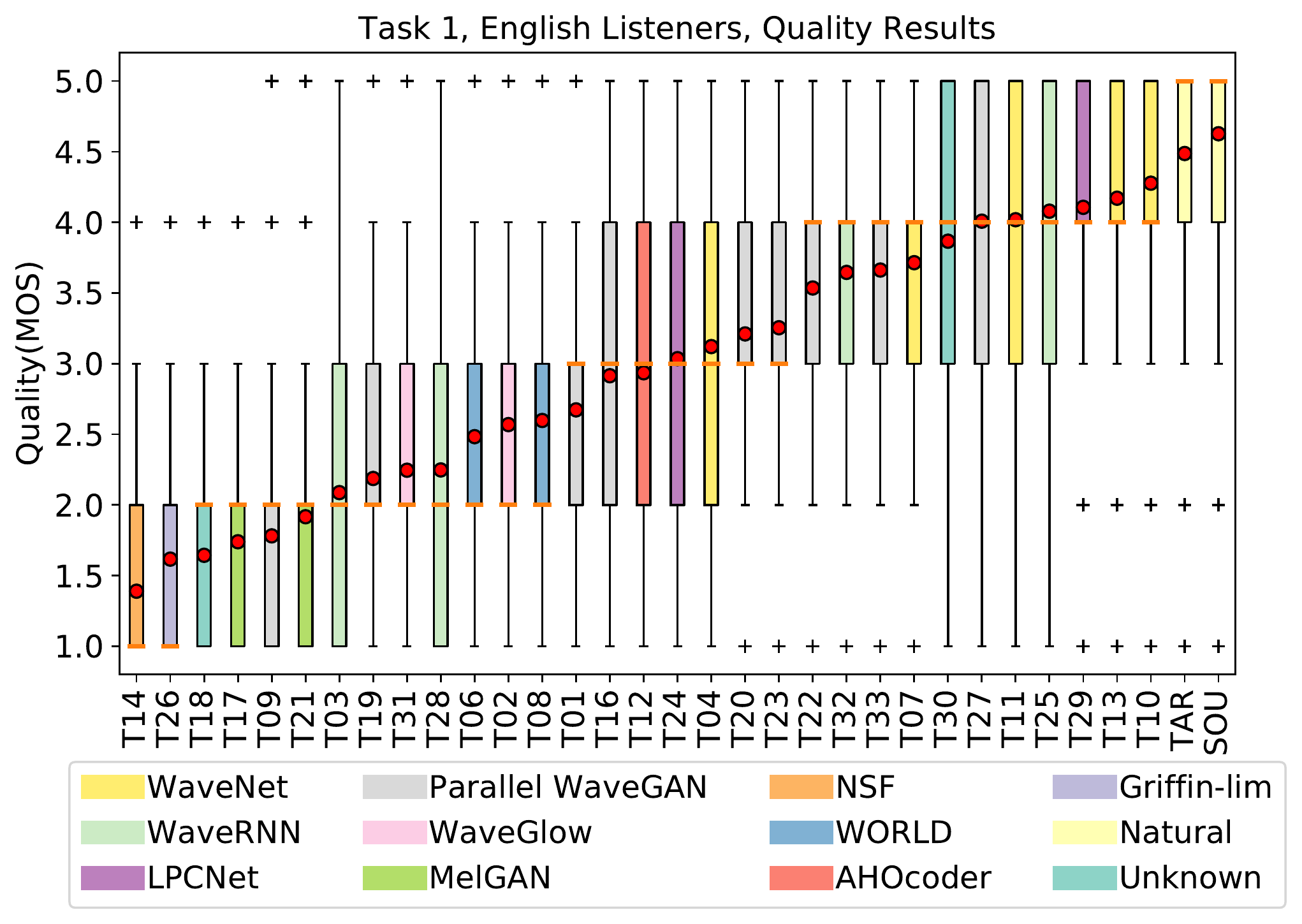} \\
\vspace{3mm}
\includegraphics[width=0.65\linewidth]{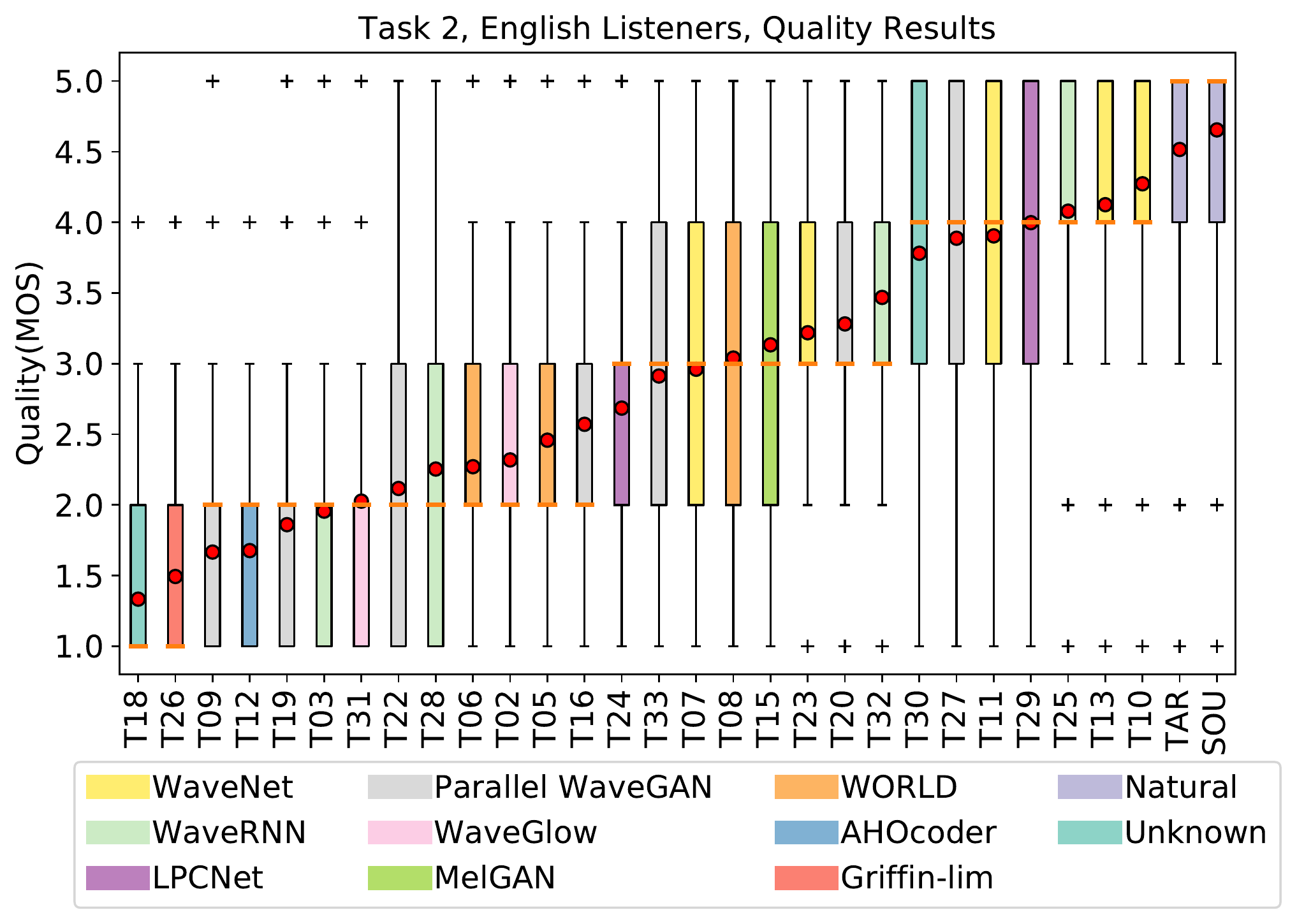} 
	\caption{Naturalness results for Task 1 (top) and Task 2 (bottom). MOS scores are arranged in accordance with their mean (red dot). Bars are colored on basis of vocoder categories.} \label{fig:en_intra_score_qua_voc}
\end{figure}

Here, we show colorization results based on the vocoder types described in Section \ref{vocoder}. The top and bottom parts of Figure \ref{fig:en_intra_score_qua_voc} show the naturalness results for Task 1 (top) and Task 2 (bottom) and the corresponding vocoder types used for each system. 

\newpage

\section{Evaluation results –– Japanese listeners --}
\label{apped:Jpn} 

\begin{figure}[h]
	\centering
	\includegraphics[width=0.6\linewidth]{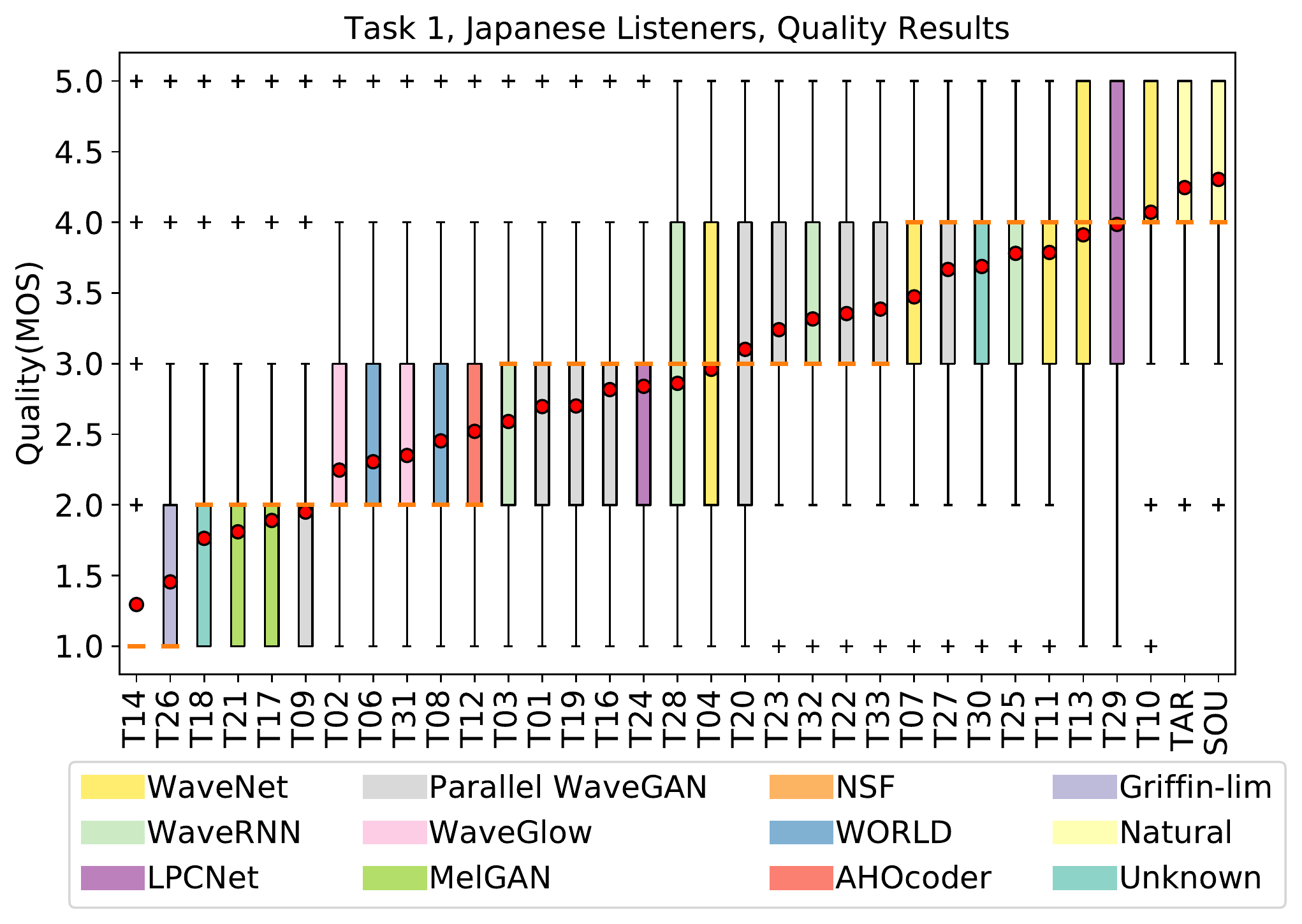} \\	\includegraphics[width=0.8\textwidth]{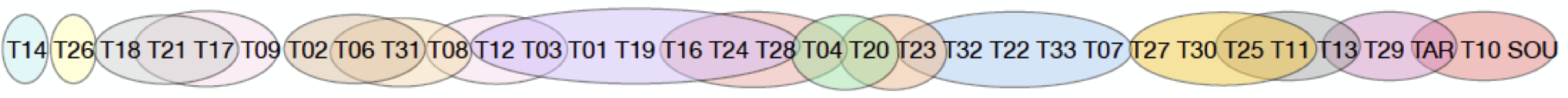} 
	\caption{Naturalness results for Task 1 and groupings of systems that did not differ significantly from each other. Bars are colored based on vocoder categories. }	\label{fig:jp_intra_score_qua}
\end{figure}

\begin{figure}[h]
	\centering
 	\vspace{-0.7cm}
	\includegraphics[trim=1cm 0 1cm 2cm,clip,width=0.65\linewidth]{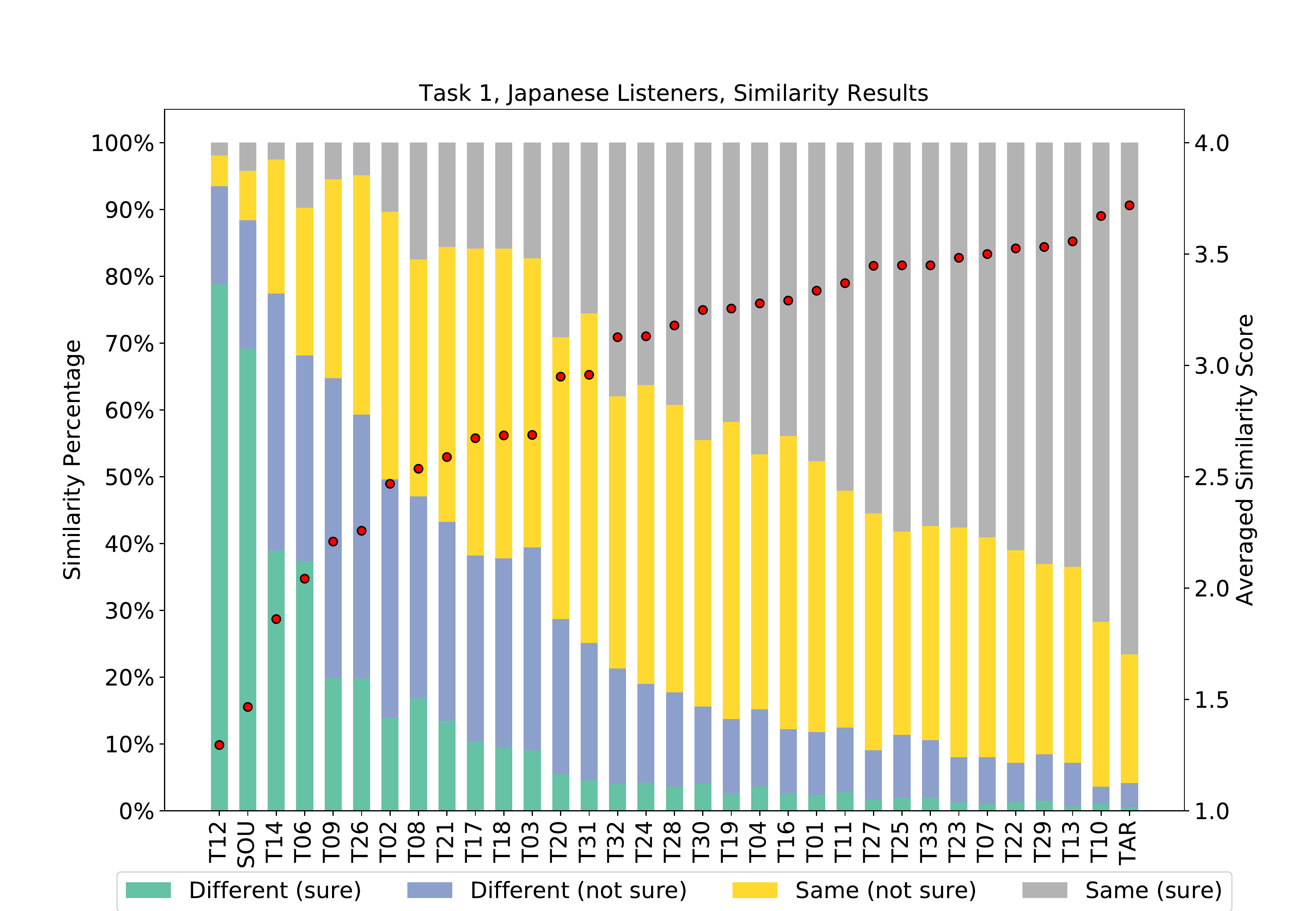} 
	\includegraphics[width=0.8\textwidth]{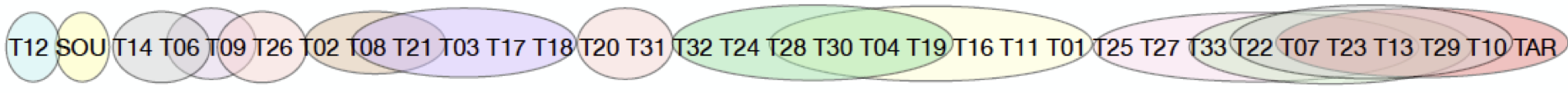} 
	\caption{Similarity results of target speaker for Task 1 and groupings of systems that did not differ significantly from each other.}\label{fig:jp_intra_score_sim}
\end{figure}

\begin{figure}[h]
	\centering
	\includegraphics[width=0.65\linewidth]{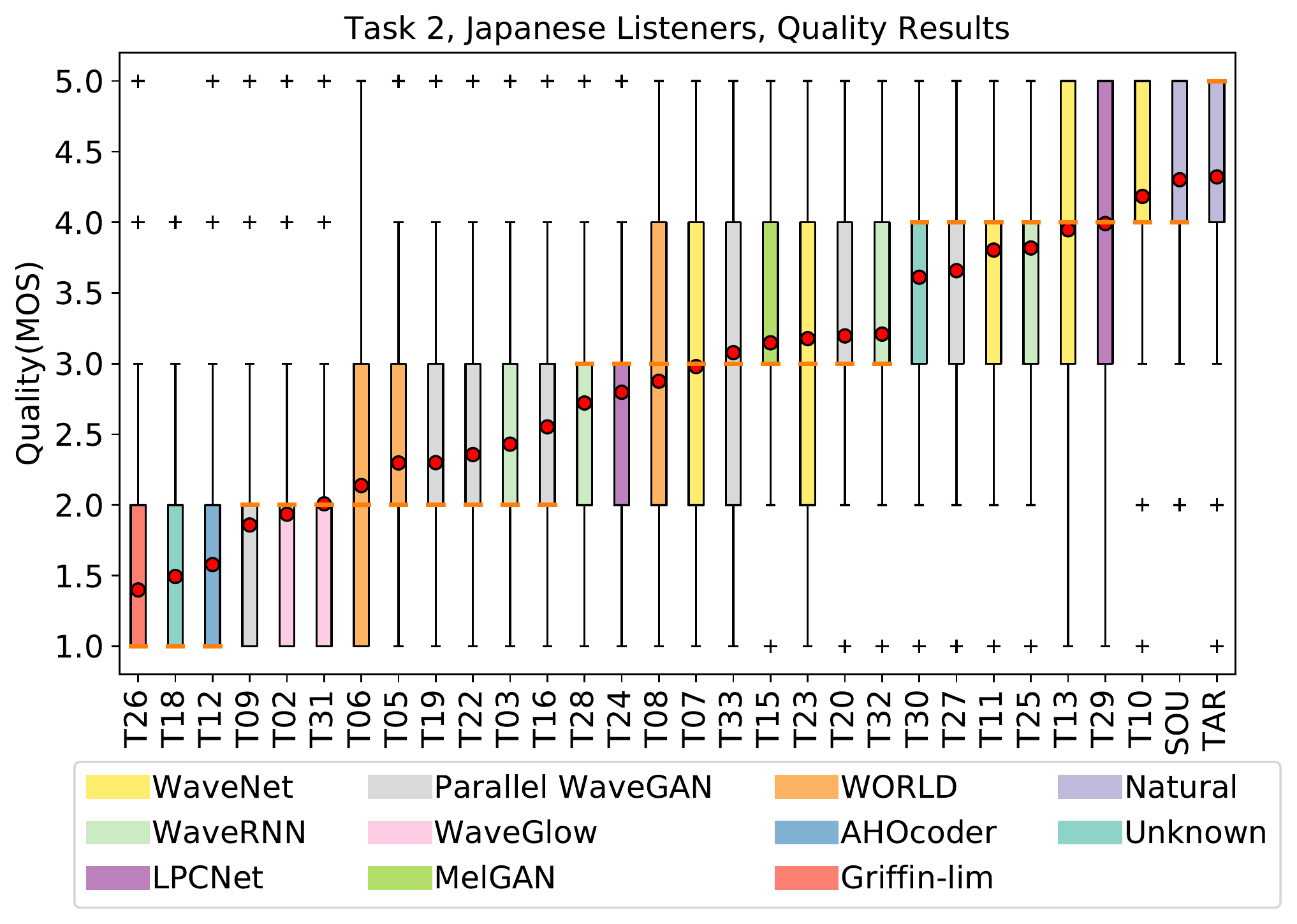} \\
	\includegraphics[width=0.8\textwidth]{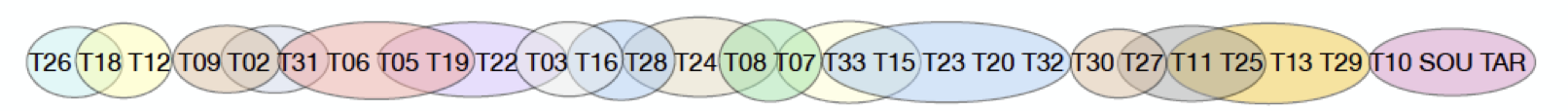} 
	\caption{Naturalness results for Task 2 and groupings of systems that did not differ significantly from each other.  \label{fig:jp_cross_score_qua}}
\end{figure}

\begin{figure}[h]
	\centering
	\includegraphics[trim=1cm 0 1cm 0,clip,width=0.7\linewidth]{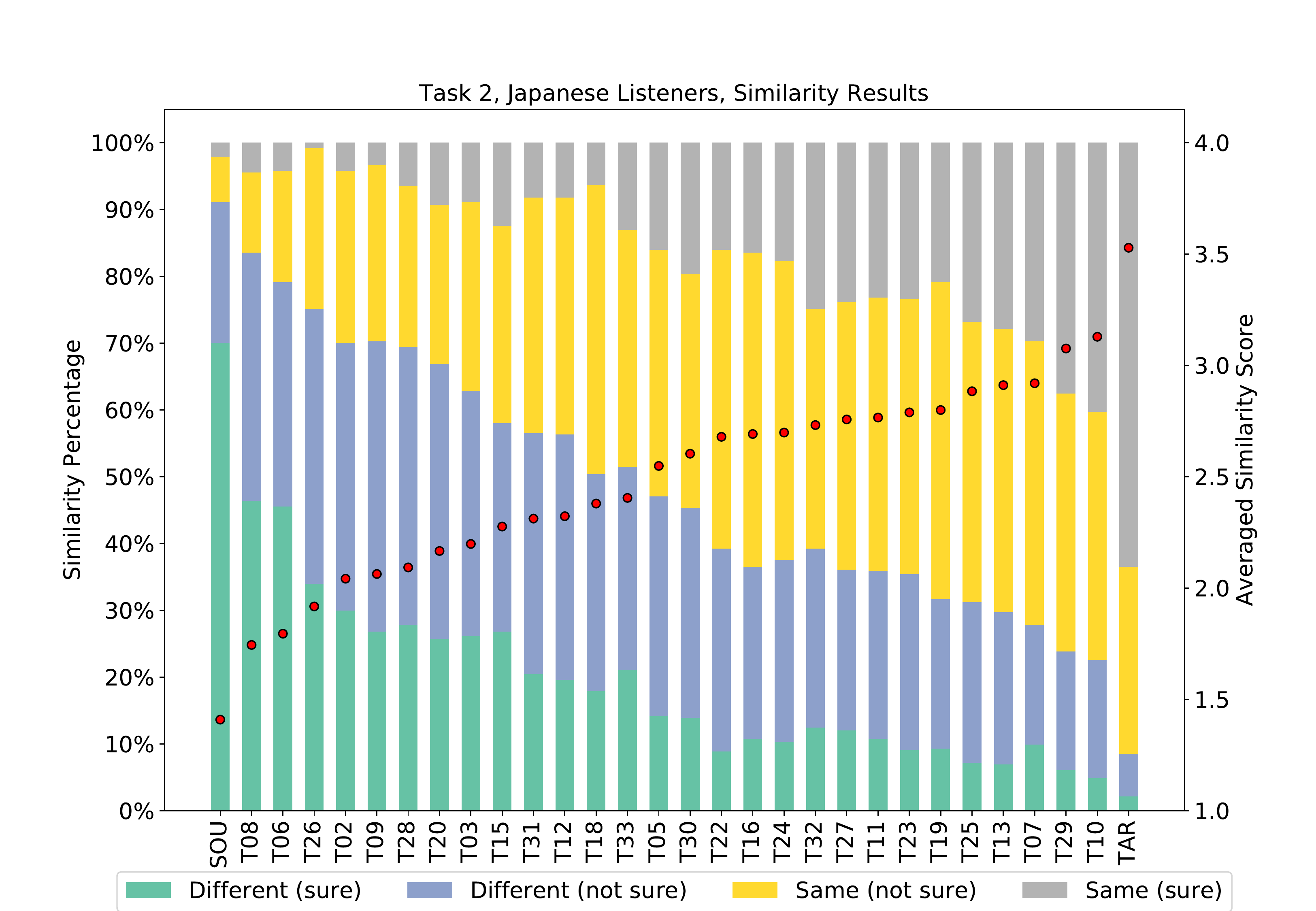} \\
	\includegraphics[width=0.8\textwidth]{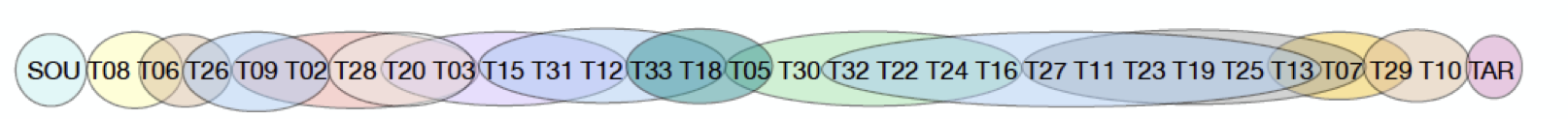} 

	\caption{Similarity results of target speaker for Task 2. Similarity scores are arranged in accordance with their mean value (red dot).	\label{fig:jp_cross_score_sim}}
\end{figure}

Here, we show the evaluation results for the Japanese listeners. The total number of unique valid listeners was 206. We used the same figure formats as for the English results described in Section \ref{main-result}. Figures \ref{fig:jp_intra_score_qua} and \ref{fig:jp_intra_score_sim} show the results of the naturalness and speaker similarity evaluation for Task 1, respectively, and Figures \ref{fig:jp_cross_score_qua} and \ref{fig:jp_cross_score_sim} show those for Task 2, respectively.

The general tendencies were the same as those of the English listeners. For instance, the speaker similarity scores for several of the systems were as high as the target speakers in Task 1, whereas all of the VC systems had lower speaker similarity scores than the target speakers in Task 2. One difference is that the naturalness of T10 was rated as good as the natural speech of the target speakers (as well as speaker similarity) in Task 1, and there were no significant differences between them. For the Japanese listeners, the converted audio produced by T10 sounded as if it were from target speakers. This was not the case for the English listeners.

\end{document}